\documentclass[twocolumn,floatfix,prb,aps,showpacs]{revtex4}
\usepackage{graphicx,amsmath,amssymb,color}
\usepackage{nicefrac}
\usepackage[titletoc,title]{appendix}

\newcommand{\be}{\begin{equation}}
\newcommand{\ee}{\end{equation}}

\newcommand{\ba}{\begin{eqnarray}}
\newcommand{\ea}{\end{eqnarray}}

\begin{document}

\title{Fractional Quantum Hall Effect in Graphene: Quantitative Comparison between Theory and Experiment}
\author{Ajit C. Balram,$^1$ Csaba T\H oke,$^2$ A. W\'ojs,$^3$ and J. K. Jain,$^1$}

\affiliation{
   $^{1}$Department of Physics, 104 Davey Lab, Pennsylvania State University, University Park, PA 16802, USA}
\affiliation{
   $^{2}$BME-MTA Exotic Quantum Phases ``Lend\"ulet" Research Group, Budapest University of Technology and Economics,
Institute of Physics, Budafoki \'ut 8., H-1111 Budapest, Hungary}
\affiliation{
   $^{3}$Department of Theoretical Physics, Wroclaw University of Technology, Wybrzeze Wyspianskiego 27, 50-370 Wroclaw, Poland}

\begin{abstract} 
The observation of extensive fractional quantum Hall states in graphene brings out the possibility of more accurate quantitative comparisons between theory and experiment than previously possible, because of the negligibility of finite width corrections. We obtain accurate phase diagram for differently spin-polarized fractional quantum Hall states, and also estimate the effect of Landau level mixing using the modified interaction pseudopotentials given in the literature. We find that the observed phase diagram is in good quantitative agreement with theory that neglects Landau level mixing, but the agreement becomes significantly worse when Landau level mixing is incorporated assuming that the corrections to the energies are linear in the Landau level mixing parameter $\lambda$. This implies that a first order perturbation theory in $\lambda$ is inadequate for the current experimental systems, for which $\lambda$ is typically on the order of or greater than one. We also test the accuracy of the composite-fermion theory and find that all lowest Landau level projection methods used in the literature are very accurate for the states of the form $n/(2n+1)$ but for the states at $n/(2n-1)$ the results are more sensitive to the projection method. An earlier prediction of an absence of spin transitions for the $n/(4n+1)$ states is confirmed by more rigorous calculations, and new predictions are made regarding spin physics for the $n/(4n-1)$ states.
\pacs{73.43.Cd, 71.10.Pm}
\end{abstract}
\maketitle

\section{Introduction}

A longstanding issue in the field of the fractional quantum Hall effect (FQHE)\cite{Tsui82} has been that the quantitative agreement between theory and experiment is less precise than what one would expect. It is possible to obtain very accurate numbers for many quantities of interest from the composite fermion (CF) theory \cite{Jain89,Jain07,Jain15}. Detailed comparisons have been carried out for activations gaps \cite{Du93}, collective mode dispersions\cite{Kukushkin09}, and spin-polarization phase transitions \cite{Jain07}. In all cases, the measured numbers are generally consistent with those predicted by theory, but the agreement is worse than that suggested by the accuracy of the theory as determined from comparisons with exact diagonalization results\cite{Dev92,Wu93,Jain07,Balram13}. It is believed that the deviation arises from corrections due to effects extraneous to the FQHE physics, such as finite quantum well width, Landau level (LL) mixing and disorder, which provide sizable corrections that are hard to deal with in a quantitative manner. 

The observation of FQHE in graphene \cite{Xu09,Bolotin09,Dean11,Feldman12,Feldman13,Amet15,Lin14} provides a unique opportunity in this context, because finite width corrections are essentially absent in graphene. Furthermore, as noted by Peterson and Nayak\cite{Peterson14}, unlike in GaAs quantum wells, in the $n=0$ LL of graphene, LL mixing does not produce any effective three body interaction (which incorporates the breaking of particle-hole symmetry), but only corrections to the pairwise interaction. One might therefore expect smaller corrections due to LL mixing in graphene than in GaAs. The FQHE in graphene may thus provide an opportunity for better quantitative comparisons between theory and experiments, and a better appreciation of our understanding of the role of LL mixing. 

The spin polarization transitions provide some of the most precise tests of the quantitative accuracy of the FQHE theory, for several reasons. First, it is a thermodynamic measurement (as opposed to the excitation energies of charged or neutral modes), and therefore is likely to be less susceptible to the presence of disorder. Second, the critical Zeeman energy where a transition between two differently spin polarized FQHE states occurs provides a direct measure of the rather small Coulomb energy differences between the two states, and thus enables a detailed and exceedingly sensitive test of our quantitative understanding of the FQHE. Finally, there is an extensive amount of experimental phenomenology associated with the spin physics. Phase transitions as a function of the Zeeman energy have been measured in various semiconductor based two-dimensional systems  \cite{Eisenstein89,Du95,Yeh99,Kukushkin99,Kukushkin00,Melinte00,Tiemann12,Liu14,Liu15,Bishop07,Padmanabhan09,Shabani10,Padmanabhan10,Gokmen10b,Liu14a, Betthausen14}, and recently also in graphene by Feldman {\em et al.}\cite{Feldman13}. 

Hoping to obtain a better comparison between theory and experiment, we have determined the spin-polarization phase diagram as accurately as possible for an ideal two-dimensional system in the $n=0$ LL. (These results apply to graphene as well as to narrow GaAs quantum wells, because in the limit of zero thickness, the $n=0$ LL wave functions are identical for the two.) We perform large scale exact diagonalization studies of various FQHE states for this purpose. This also allows us to determine the quantitative accuracy of the CF theory\cite{Jain89,Jain07,Lopez91,Halperin93}. We also estimate the effect of LL mixing by modifying the pseudopotentials according to Ref.~\cite{Peterson14}. Our conclusions, briefly, are as follows.

\begin{figure*}
\begin{center}
\includegraphics[width=0.8\textwidth,height=0.45\textwidth]{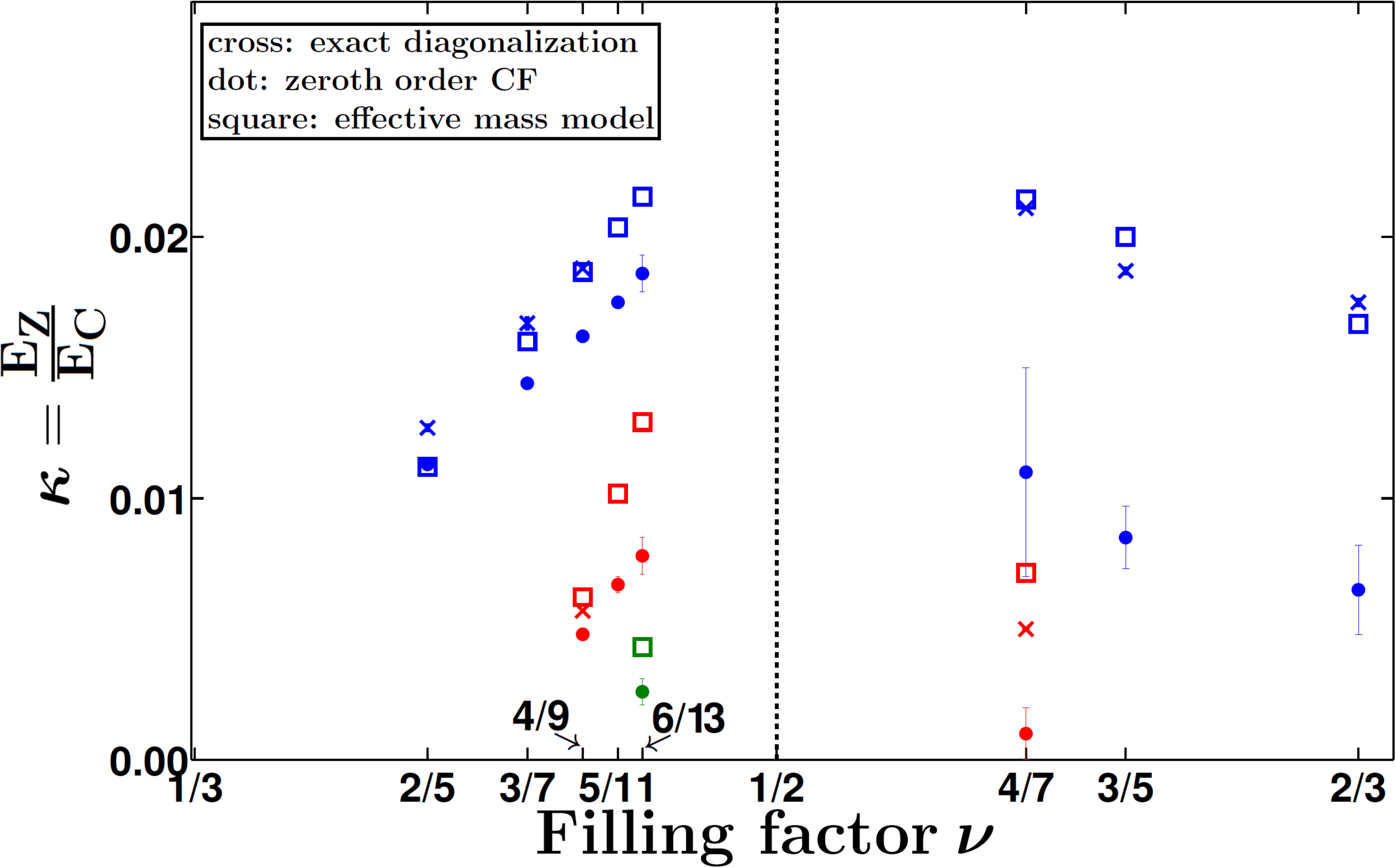}
\caption{(Color online) Spin phase diagram showing the critical Zeeman energies for transitions of the fractional quantum Hall states at the filling factor $\nu=n/(2n\pm 1)$. We define $\kappa=E_{\rm Z}/E_{\rm C}$, where $E_{\rm Z}$ is the Zeeman splitting and $E_{\rm C}=e^2/\epsilon l$ is the Coulomb energy scale. The crosses show the results from exact diagonalization (assuming zero thickness), the dots from JK wave functions, and squares are fits from the free-CF model (see text). The blue, red and green colors indicate transition from $(n,0)\rightarrow (n-1,1)$, $(n-1,1)\rightarrow (n-2,2)$ and $(n-2,2)\rightarrow (n-3,3)$ respectively. We note that the exact diagonalization results for the spin transitions at $\nu=4/7$ and $\nu=4/9$ are obtained using results extrapolated from only two finite systems, and may therefore be less accurate.}
\label{fig:spin_phase}
\end{center}
\end{figure*}

From exact diagonalization calculations, we have obtained essentially the exact theoretical phase diagram of the spin polarization of FQHE states without allowing for LL mixing (see blue crosses in Fig.~\ref{fig:spin_phase}). We find that it is in excellent agreement with the experimental phase diagram in graphene (see dots in Fig.~\ref{fig:spin_phase}). The agreement between theory and experiment becomes significantly worse when LL mixing is included to linear order in  the LL mixing parameter $\lambda$ using the pseudopotentials of Ref.~\onlinecite{Peterson14}. This suggests that the current experimental systems are outside the linear regime, which is not surprising given that $\lambda$ is typically of order 1 and sometimes much larger (e.g. 2.2 for suspended graphene). This is relevant to the issue of whether the Pfaffian or the anti-Pfaffian state\cite{Moore91,Levin07,Lee07} at $\nu=1/2$ in the second LL is selected by LL mixing, which has received much attention recently\cite{Wojs10,Rezayi11,Peterson14,Zaletel15,Pakrouski15,Tyler15}. 

The allowed spin polarizations for various FQHE states at $n/(2n\pm 1)$ and their energy ordering are correctly predicted by the CF theory. The mean-field model in which composite fermions are treated as free particles at an effective magnetic field\cite{Halperin93,Park98,Jain07} remains satisfactory, and we obtain a precise estimation for the CF mass. We also carry out quantitative tests of the CF theory, which are extremely precise because the critical Zeeman energies depend on very small energy {\em differences} between different states, and even a slight error in the energy can lead to large corrections in the critical Zeeman energies. For the fractions $\nu=n/(2n+1)$, calculations based on Jain's wave functions\cite{Jain89} predict the critical Zeeman energies with 15\% accuracy, which can be further improved by the method of CF diagonalization that incorporates $\Lambda$ level ($\Lambda$L) mixing, where a $\Lambda$L refers to a CF Landau level in an effective magnetic field. For the states at $\nu=n/(2n-1)$, the results depend sensitively on the lowest Landau level (LLL) projection method. In particular, the treatment with Jain-Kamilla (JK) projection\cite{Jain97,Jain97b,Jain07} underestimates the critical Zeeman energies by a factor of 2-3 (i.e., relatively overestimates the energies of the non-fully spin polarized states). This is not an intrinsic deficiency of the CF theory but rather a technical issue, as can be seen from the fact that the ``hard-core projection" introduced previously in Ref.~\onlinecite{Wu93} produces very accurate states for $\nu=n/(2n-1)$. (This projection is presently not amenable to calculations for large systems.) 

For completeness, we have also considered the states of composite fermions with four vortices attached. A previous prediction\cite{Park99} of the absence of spin transitions for the $n/(4n+1)$ states is confirmed by more rigorous calculations. Here the states remain spontaneously spin polarized even in the absence of a Zeeman energy. Spin transitions are, however, possible for the $^4$CF FQHE state at $\nu=n/(4n-1)$, and we estimate the critical Zeeman energies for the prominent transition.

\section{FQHE in graphene}
The physics of graphene differs from that of GaAs in two important aspects: the dispersion of electrons is linear and there are four Dirac cones.  The linear dispersion of Dirac fermions leads to LLs which have a cyclotron energy of sgn$(n)\sqrt{2|n|}\hbar v_{F}/\ell$, where $v_{F}$ is the Fermi velocity and $\ell$ is the magnetic length and $n$ is any integer\cite{Neto09,Goerbig11}. (This is to be contrasted from GaAs or other conventional semiconductors wherein electrons have a parabolic dispersion and LLs which have a cyclotron energy given by $\hbar\omega_c (n+1/2)$, where $n$ is a non-negative integer and $\omega_c$ is the cyclotron energy. ) Also, the LL wave functions are two component wave functions, corresponding to two sublattices of graphene. In general this leads to different interaction pseudopotentials\cite{Apalkov06,Nomura06,Toke06} than in systems with parabolic dispersion. However, it turns out that the electron wave functions as well as the interaction pseudopotentials in the $n=0$ LL of graphene are identical to those in the LLL of GaAs. 

The second difference is that each LL of graphene has four bands, which arise from the valley and spin degrees of freedom, while GaAs has two bands from the spin degree of freedom. Within each band, the physics of FQHE in the $n=0$ LL is identical in the two systems, apart from corrections due to finite width and LL mixing. Much work has been done toward understanding the origin of the lifting of various degeneracies\cite{Morpurgo06,Aleiner07,Semenoff12,Roy14}, but we will assume below that all bands are well separated.  This assumption is justified for the experiments with which we compare our results. 

We refer the reader interested in further information to an extensive literature on the physics of Dirac fermions in a magnetic field\cite{Neto09,Goerbig11}.

\section{Phase diagram of spinful CF states} 

A good qualitative and semiquantitative theoretical understanding of these transitions has been obtained in terms of integer or fractional quantum Hall effect of spinful composite fermions \cite{Wu93,Park98,Park98b,Park99,Park01,Chang03b,Murthy03,Wojs07,Davenport12,Mukherjee14,Balram15d,Balram15}, which successfully predicts the allowed spin polarizations at all of these filling factors and also provides an estimate of the the critical Zeeman energy where transitions between them occur. While these quantitative estimates are a good zeroth order approximation, their accuracy has not been carefully evaluated in the past.

The FQHE state at $\nu=n/(2pn\pm 1)$ maps into integer quantum Hall effect (IQHE) state of composite fermions with $n$ filled $\Lambda$ levels, where a composite fermion is defined as a bound state of an electron and $2p$ vortices. For spinful composite fermions, the CF filling is written as $n=n_\uparrow+n_\downarrow$, where $n_\uparrow$ and $n_\downarrow$ are the number of filled spin-up and spin-down $\Lambda$Ls. The different states will be denoted as $(n_\uparrow,n_\downarrow)$, and we will use the convention $n_\uparrow\geq n_\downarrow$ without loss of generality. One can list all possible states and their spin polarization $\gamma=(n_\uparrow-n_\downarrow)/(n_\uparrow+n_\downarrow)$. For example, for 2/5 and 2/3, which both map into $n=2$ of composite fermions, we have a fully spin polarized state $(2,0)$ (with $\gamma=1$) and a spin singlet state $(1,1)$ (with $\gamma=0$). To take another example, 6/13 and 6/11 map into $n=6$, where we have four possible states $(6,0)$, $(5,1)$, $(4,2)$ and $(3,3)$, with $\gamma=1$, 2/3, 1/3, and 0, respectively. One expects one transition at 2/5 and 2/3 and three at 6/13 and 6/11. The possible states and spin polarizations of other fractions can be similarly enumerated.

The CF theory also identifies the flux values where these states occur in the spherical geometry \cite{Jain07}. These are the flux values at which our calculations below are carried out. All our calculations are performed in the spherical geometry, where the ground states have total orbital angular momentum $L=0$ and a total spin $S$ that can be ascertained by the CF theory. We will assume the ideal limit of zero thickness and neglect disorder. In quoting the energies below, we include the electron-background and background-background interaction. The density for a finite system (in the spherical geometry) depends on the number of particles $N$ and is slightly different from its thermodynamic limit. To eliminate this effect we use the density corrected energy \cite{Morf86}  $E_{N}^{'}=(2Q\nu/N)^{1/2}E_{N}$ for extrapolation to the thermodynamic limit $N^{-1}\rightarrow 0$, where the integer $2Q$ is the magnetic flux (in units of $\phi_0=hc/e$) to which the electrons are subjected. All energies quoted below are the thermodynamic limits of the per particle density corrected energies $\text{lim}_{N\rightarrow\infty} E_{N}^{'}/N$.

To avoid clutter, we give only give the spin-polarization phase diagram in the main text. All of the individual numbers as well as extrapolations are given in various tables and figures in Appendix \ref{appendix1}.

\subsection{Exact diagonalization} We first obtain the extrapolated values of energies of the variously polarized states at fractions of the form $n/(2n\pm 1)$. These include the largest systems for which exact diagonalization can currently be performed (see Appendix \ref{appendix1} for Hilbert space dimensions). For filling factors $4/7$ and $4/9$, the extrapolated values are obtained with only two points and thus must be treated with caution, but we have chosen to include them because linear extrapolation in $1/N$ has been found to be quite accurate for other systems for which several points are available. Once we have the energies, we obtain the critical Zeeman energies by setting the energy difference of the two successive states $(n_\uparrow,n_\downarrow)$ and $(n_\uparrow-1,n_\downarrow+1)$ to zero:
\be
[\delta_{(n_\uparrow,n_\downarrow)}-\delta_{(n_\uparrow-1,n_\downarrow+1)}]\;{e^2\over \epsilon \ell}-{1\over n} E_{\rm Z}=0
\ee
where we have used $\delta_{(n_\uparrow,n_\downarrow)}$ to denote the thermodynamic limit of the per particle density-corrected, background-subtracted Coulomb interaction energy of the state $(n_{\uparrow},n_{\downarrow})$. This gives 
\be
\kappa:= {E_{\rm Z}\over  e^2/( \epsilon \ell)} = n [\delta_{(n_\uparrow,n_\downarrow)}-\delta_{(n_\uparrow-1,n_\downarrow+1)}]
\ee
where $\epsilon$ is the dielectric constant of the host material and $\ell=\sqrt{\hbar c/eB}$ is the magnetic length. The resulting critical energies are shown by the blue crosses in Fig.~\ref{fig:spin_phase}.

\subsection{Free CF model} 

We ask to what extent the results may be interpreted in a model that treats composite fermions as free particles with an effective mass \cite{Halperin93}, which has been used routinely to analyze the experimental data \cite{Du95,Melinte00,Feldman13}. The interaction energy between electrons is modeled in terms of the CF cyclotron energy, defined as 
\be
\hbar\omega_c^*=\hbar {eB^*\over m_p^* c}=\hbar {eB\over (2pn\pm 1) m_p^* c}\equiv {\alpha \over 2pn\pm 1} \;{e^2\over \epsilon \ell}
\ee
where $m_p^*$ is the CF mass. (The subscript is to remind us that this mass is the ``polarization mass" of composite fermions,\cite{Park98} which is the relevant mass for the spin-polarization phase transitions. It is to be distinguished from the mass defined from the activation gap \cite{Halperin93}.) The CF mass is often quoted in units of the electron mass in vacuum, $m_e$:
\be
{m_p^*\over m_e}={1\over \alpha} \; {\hbar \Omega_c\over e^2/\epsilon \ell}
\ee
where $\Omega_c=eB/m_ec$ is the cyclotron frequency of electron in vacuum. The CF mass behaves as $m_p^*\sim \sqrt{B}$ and for the parameters of GaAs, we have 
\be m_p^*/m_e=(0.026/\alpha)\sqrt{B[T]}.
\ee

An immediate qualitative prediction of the free-CF model is that the interaction energies increase with the degree of spin polarization. This has been found to be the case for all states of the form $n/(2n\pm 1)$. [As seen below, this is not the case for $n/(4n+1)$.] At a more quantitative level, the free-CF model predicts that the critical Zeeman energy for the transition between $(n_\uparrow,n_\downarrow)$ and $(n_\uparrow-1,n_\downarrow+1)$ is given by
\be
{1\over n}[n_\uparrow-n_\downarrow-1]\hbar\omega_c^*-{1\over n} E_{\rm Z}=0
\ee
which gives
\be
\kappa=\alpha\; {n_\uparrow-n_\downarrow-1 \over 2pn\pm 1}
\ee
We have found that the best fit for the critical Zeeman energies calculated here is provided by $\alpha_{n/(2n+1)}=0.056$ and $\alpha_{n/(2n-1)}=0.050$. A single value of $\alpha$ gives a slightly less satisfactory fit, implying a weak filling factor dependence for the CF mass. Nonetheless, the free-CF model works reasonably well. We note that the CF polarization mass $m_p^*$ for the reverse flux attached states is about 10\% higher than that for parallel flux attached states.

\subsection{Microscopic theory} 

The wave function for the ($n_{\uparrow},n_{\downarrow}$) state at $n/(2pn\pm 1)$ state is given by 
\be
\Psi_{\frac{n}{2pn+1}}={\cal P}_{\rm LLL} \Phi_{n} J^{2p}={\cal P}_{\rm LLL} \Phi_{n_{\uparrow}}\Phi_{n_{\downarrow}} J^{2p}
\label{paral-flux}
\ee
and 
\be
\Psi_{\frac{n}{2pn-1}}= {\cal P}_{\rm LLL} [\Phi_{n}]^{*} J^{2p}= {\cal P}_{\rm LLL} [\Phi_{n_{\uparrow}}\Phi_{n_{\downarrow}}]^{*} J^{2p}
\label{rev-flux}
\ee
where 
\be
\label{Jastrow}
J= \prod_{1\leq j<k \leq N}(z_j-z_k)
\ee 
is the Jastrow factor, $z_i$ is the coordinate of the $i^{\text{th}}$ electron, $\Phi_{n_\uparrow}$ ($\Phi_{n_\downarrow}$) is the Slater determinant wave function for $n_{\uparrow}$ ($n_{\downarrow}$) filled LLs of electrons, and ${\cal P}_{\rm LLL}$ denotes LLL projection. In this section we shall restrict ourselves to the case of $p=1$ and consider higher values of $p$ in the subsequent section. Three schemes have been employed for LLL projection, which result in slightly different LLL projected wave functions. (i) ``Direct projection" will refer to the method considered in Refs. \onlinecite{Dev92} and \onlinecite{Wu93}, wherein the product wave function is expanded into Slater determinant basis and only the part strictly in the LLL is retained. This method can be implemented for relatively small systems (ten particles or fewer). (ii) In the ``hard-core projection" of Ref.~\onlinecite{Wu93}, one writes the wave function as 
\be
\Psi^{\rm hard-core}_{\frac{n}{2n+1}}= J\; {\cal P}_{\rm LLL} \Phi_{n} J
\ee
\be
\Psi^{\rm hard-core}_{\frac{n}{2n-1}}= J\; {\cal P}_{\rm LLL} [\Phi_{n}]^{*} J
\ee
As the name implies, this method explicitly builds correlations such that the wave function vanishes even when particles of opposite spin coincide. This method also relies on expansion into Slater determinant basis and can be implemented only for small systems. (iii) The most widely used projection method is the so-called JK projection, which has the advantage that it does not require expansion into Slater determinant basis and thus can be evaluated for very large systems. This method has been used extensively to make quantitative predictions for various quantities. The details of the JK projection method have been outlined in the literature \cite{Jain97,Jain97b,Jain07} and will not be repeated here. We will refer to the resulting wave functions as JK wave functions, to distinguish them from wave functions obtained by other projections \cite{Dev92,Dev92a,Wu93}. 

The results from the JK projection are also shown in Fig.~\ref{fig:spin_phase}. For the ``parallel flux-attached" states at $n/(2n+1)$, the JK wave functions underestimate the critical Zeeman energies  (also given previously in Ref.~ \onlinecite{Park98}) by $\sim$ 15\%. As a result, the $\alpha\approx 0.056$ is lower, and the CF effective mass is higher, by $\sim$15 \% compared to the values from exact diagonalization. For the ``reverse-flux attached" states at $n/(2n-1)$, the JK wave functions underestimate the critical Zeeman energies by a factor of two to three. These results bring out the limitations of the JK projection method for the reverse-flux attached states.

From our current study as well as the previous results\cite{Dev92,Wu93} (reproduced in Table \ref{tab_proj}) we find that: the hard-core projection produces very accurate energies for fully as well as non-fully spin polarized states at $n/(2n+1)$ and $n/(2n-1)$; the JK and the Direct projections produce accurate results for fully polarized states at $n/(2n+1)$ and $n/(2n-1)$ and also for non-fully polarized states at $n/(2n+1)$; the JK and Direct projections are somewhat less accurate for non-fully polarized states at $n/(2n-1)$.  It is easy to see why the JK / Direct projection underestimates the critical Zeeman energies: its deficiency is that it does not eliminate configurations in which spin-up and spin-down particles are coincident, and thus overestimates the energies of non-fully spin polarized states by a larger amount than for the fully spin polarized states, thus resulting in an underestimation of the critical Zeeman energies. It is unclear why the JK and Direct projections work better for the non-fully polarized states at $n/(2n+1)$ than those at $n/(2n-1)$.

\begin{table}
\begin{tabular}{|c|c|c|c|c|c|c|c|c|c|c|}
\hline
\multicolumn{1}{|c|}{$\nu$} & \multicolumn{3}{|c|}{system} & \multicolumn{3}{|c|}{\% error in the Coulomb interaction energy} \\ \hline
	    & N & 2Q &  state	& JK	&  Direct	&  hard-core	\\ \hline

2/3	    & 8 & 11 &  (1,1)	&  0.75	&	-					&	0.04 (Ref.~[\onlinecite{Wu93}])		\\ \hline 
2/3	    & 8 & 12 &  (2,0)	&  0.13	&	0.02 (Ref.~[\onlinecite{Wu93}])		&	-	\\ \hline 
2/5	    & 8 & 16 &  (2,0)	&  0.01	&	0.01 (Ref.~[\onlinecite{Dev92a}])	&	-	\\ \hline
3/5	    & 8 & 13 &  (2,1)	&  0.29	&	-					&	0.05	(Ref.~[\onlinecite{Wu93}])	\\ \hline  
3/5	    & 9 & 16 &  (3,0)	&  0.03	&	0.01 (Ref.~[\onlinecite{Wu93}])		&	-	\\ \hline  
\end{tabular}
\caption {Percent error in the Coulomb interaction energy of the CF wave function obtained from different projection methods for several finite systems for which the exact results are known. The last two columns give results from Direct projection for fully spin polarized states and from hard-core projection for the non-fully spin polarized state. (For the fully spin polarized states, Direct and hard-core projections give essentially the same energies\cite{Wu93}.) The systems correspond to $N$ particles on the surface of a sphere subjected to a total flux of $2Q hc/e$. The filling factors and the state are shown.} 
\label{tab_proj} 
\end{table}

\begin{figure}
\begin{center}
\includegraphics[width=0.45\textwidth]{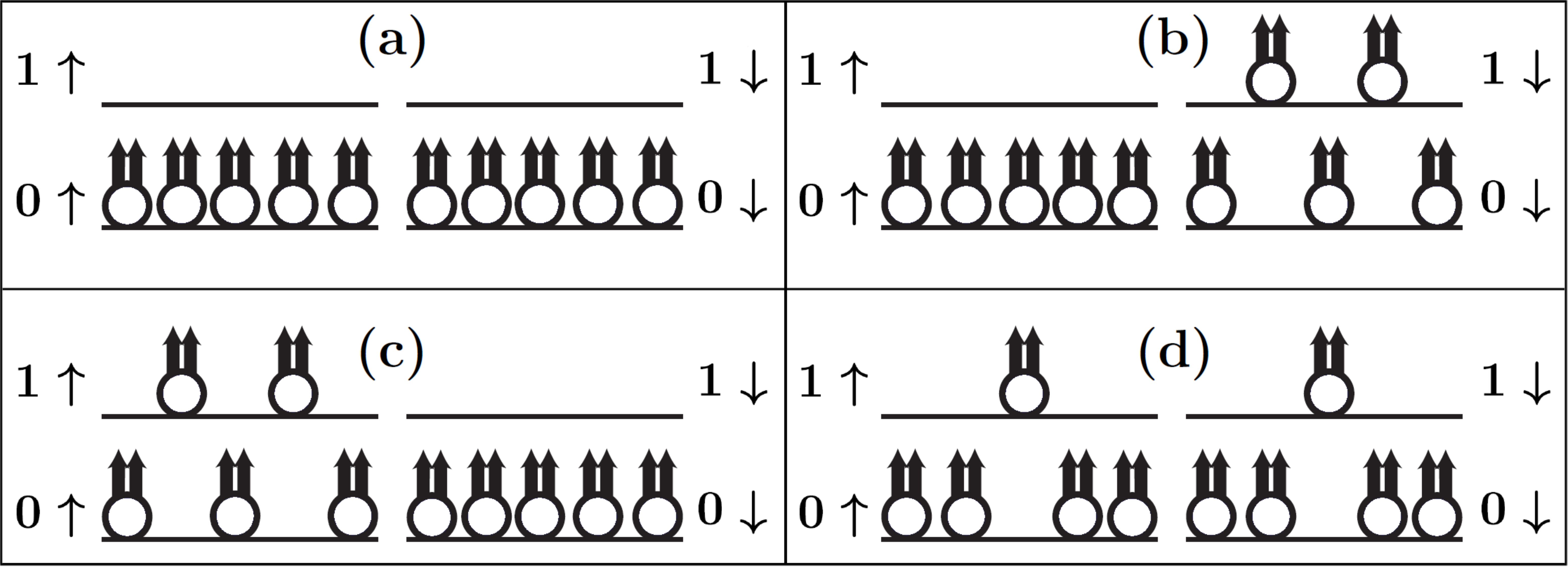}
\caption{Schematic representation of the CF basis functions used for CF diagonalization study of the $2/(4p+1)$ spin-singlet FQHE state. Panel (a) shows the ``unperturbed" state. At the first order approximation, CF diagonalization allows hybridization of the state in (a) with the states in (b), (c) and (d) to obtain a new ground state with lower energy than the unperturbed state in (a). Successive mixing with higher and higher excitations produces better approximations to the ground state.}
\label{fig:n_2_ss}
\end{center}
\end{figure}

It is in principle possible to improve the accuracy of the results within the JK projection scheme by the method of composite fermion diagonalization (CFD) \cite{Mandal02}, in which one can obtain more accurate energies by allowing some $\Lambda$ level mixing. ($\Lambda$L mixing is to be distinguished from LL mixing.) We allow $\Lambda$ level mixing by including CF excitons in the basis. A CF exciton is defined as a pair of CF particle and CF hole, where a CF particle is a CF in the lowest unoccupied $\Lambda$ level and a CF hole is a missing CF from the top most occupied $\Lambda$ level. Note that a single CF exciton does not change the ground state energy. This is easy to see since a CF hole carries an orbital angular momentum one smaller than the CF particle with which it forms the exciton. Therefore the smallest angular momentum that a CF exciton can have is $1$, thus precluding its admixture with the ground state. Therefore, we need a minimum of two CF excitons to improve the ground state. For example in Fig. \ref{fig:n_2_ss} we show the excitations we considered to improve the $\nu=2/(4p+1)$ spin-singlet state. The Hilbert space grows very quickly with the number of excitons included in the basis for CFD, so we restrict ourselves to at most two excitons. Among the wave functions shown above, the fully polarized ones are extremely accurate, so this procedure of including two CF excitons in the basis of CFD only marginally improves the ground state energy of the fully polarized state. However for the unpolarized states the improvement is substantial, evidenced by the fact that for the spin-singlet states the energies improve by around $10\%$. The method of CFD produces a critical Zeeman energy that is within $\sim$3\% of the exact value. We have carried out similar first order CFD calculations for many of the states considered in this work. These results are shown in the Appendix \ref{appendix1} and are labeled as ``CFD''. We have found CF diagonalization to be impractical for the reverse-flux attached $n/(2pn-1)$ states for technical reasons. (For states involving reverse flux attachment, the projected state is obtained as an alternating sum of elementary symmetric polynomials of high degree; see Ref.~\onlinecite{Davenport12}. To avoid the loss of significant digits we have to use software emulated multiple precision floating point numbers. The number of Monte Carlo steps to obtain the overlap and interaction matrices for CFD with reasonable accuracy is beyond our reach.)

\section{Comparison with experiments in graphene}

\begin{figure*}
\begin{center}
\includegraphics[width=0.8\textwidth,height=0.47\textwidth]{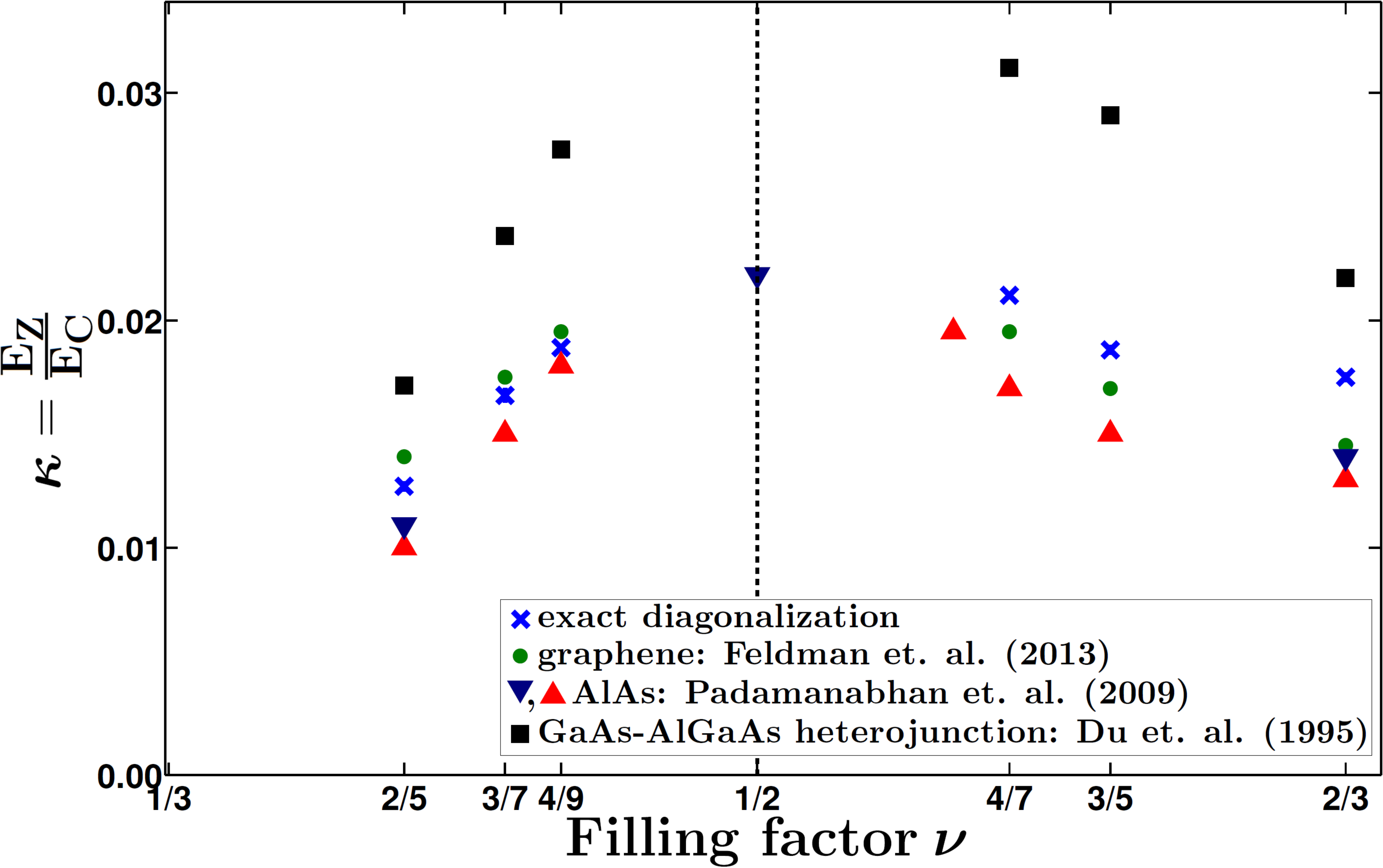}
\caption{(Color online) This figure compares the theoretical phase diagram with the experimentally measured one in graphene (taken from Feldman {\em et al.} \cite{Feldman13}, shown by green dots), GaAs (taken from Du {\em et al.} \cite{Du95}, shown by black squares) and AlAs systems (taken from Padmanabhan {\em et al.} \cite{Padmanabhan09}, shown as red or blue triangles). The theoretical values (blue crosses) are from exact diagonalization at zero thickness and zero LL mixing. (We note that the blue crosses for the spin transitions at $\nu=4/7$ and $\nu=4/9$ are obtained using results extrapolated from only two finite systems.) For AlAs there is data available at two different densities: the blue downward triangle corresponds to density of $5.5\times 10^{11}$ cm$^{-2}$ while the red upward triangle corresponds to density of $5.0\times 10^{11}$ cm$^{-2}$.}
\label{fig:spin_phase_theory_expt}
\end{center}
\end{figure*}

Fig.~\ref{fig:spin_phase_theory_expt} shows a comparison between the critical values of $\kappa$ obtained from exact diagonalization (blue crosses) with the measured ones. The results for graphene are taken from Feldman {\em et al.}\cite{Feldman13}. The value of $\kappa$ depends on the product $\epsilon g$ where $\epsilon$ is the dielectric function and $g$ is the Land\'e g factor, for which we have taken the values $\epsilon =3.0$ and $g=2.0$ for graphene in the figure. An excellent agreement between the measured and the theoretical values is evident. We stress, however, that we have not yet included effects of LL mixing, which is discussed below. 

For completeness, Fig.~\ref{fig:spin_phase_theory_expt} also shows critical Zeeman energies measured in other systems, taken from Padmanabhan et al. (AlAs quantum well) \cite{Padmanabhan09} and Du et. al \cite{Du95} (GaAs-AlGaAs heterojunction). In all cases, the critical values of $\kappa$ as well as their filling factor dependences are roughly consistent with theory. The difference between the critical values of $\kappa$ in graphene and in heterojunction samples is somewhat surprising, because the heterojunction samples also correspond to a very small thicknesses. We believe that part of the difference might result from the fact that Feldman {\em et al.} varied $\kappa$ by changing the density, whereas Du {\em et al.} \cite{Du95} accomplished that by tilting the magnetic field. We believe that high parallel fields in the latter experiments may cause additional corrections which have not been considered.

\subsection{corrections due to LL mixing in graphene}

We next come to the effect of LL mixing. The strength of LL mixing is measured by a parameter $\lambda$, which is defined as the ratio of the Coulomb to the cyclotron energy\cite{Peterson13}. For massive electrons (e.g., in GaAs) $\lambda=(e^2/\epsilon \ell)/\hbar\Omega_c$, where $\Omega_c=eB/m_{\rm b}c$ is the cyclotron frequency, with $m_{\rm b}$ being the electron band mass. For parameters appropriate for GaAs, namely $\epsilon\approx12.5$ and $m_{\rm b}\approx0.067\,m_e$, we have $\lambda\approx2.52/\sqrt{B[{\rm T}]}$, which depends on the magnetic field and falls in the range $\lambda=0.5-1.3$ for the experimentally relevant fields $B=4-25$~T. For massless Dirac electrons in graphene, $\lambda=(e^2/\epsilon \ell)/(\hbar v_{\rm F}/\ell)=e^2/(\hbar\epsilon v_{\rm F})$ is essentially the graphene fine-structure constant, independent of magnetic field. One obtains $\lambda\approx2.2$ for suspended graphene, $\lambda\approx0.9$ for graphene on SiO$_2$, $\lambda=0.5-0.8$ for graphene on BN\cite{DasSarma11,Peterson13}.

We estimate the corrections due to LL mixing in the following manner. The effect of LL mixing can be incorporated into a LLL problem by modifying the interaction, by adding a ``correction" term $V_{\rm corr}$ that contains two-, three- and higher body interaction terms. The corrections to the first few pseudopotentials of $V_{\rm corr}$ have been evaluated in the literature\cite{Bishara09,Peterson13,Peterson14} in a perturbative treatment to linear order in $\lambda$. We estimate the correction to the ground state energies by evaluating the expectation value of $V_{\rm corr}$ with respect to the unperturbed ground state. We note that for the $n=0$ graphene LL, no three body terms are induced. This method is expected to be valid for sufficiently small values of $\lambda$.

The technical details are as follows. (We include three-body interaction for completeness although it is not included in our calculations below.) For each FQHE state of interest, we evaluate the probability of occupation of various pair and triplet states, from which it is straightforward to evaluate the correction to the energy using the two- and three-body pseudopotentials of $V_{\rm corr}$. To obtain the occupation amplitudes, we have computed by exact diagonalization a sequence of finite size Coulomb ground state vectors, labeled by $N$ and $2Q=N/\nu-\sigma$ (where $\sigma$ is the ``shift'' dependent on $\nu$ and $\gamma$). 
For each vector, we then calculated the series of pair and triplet ($K=2$ and 3) amplitudes $P_{\nu,\gamma;N}^{(K)}(s,m)$ for all possible pair and triplet spins ($s=0$ and 1 for $K=2$; $s=1/2$ and $3/2$ for $K=3$) and the leading relative angular momenta (even $m=0$, 2, \dots, 8 for $K=2$ and $s=0$; odd $m=1$, 3, \dots, 9 for $K=2$ and $s=1$; $m=1$, 2, 3 for $K=3$ and $s=1/2$; $m=3$, 5, 6, 7, 8, 9 for $K=3$ and $s=3/2$), as expectation values of the corresponding model $K=2$ and 3 body pseudopotentials $V^{(K)}(s,m)$. Owing to their regular size dependence, each amplitude was then reliably extrapolated by a linear regression as a function of $1/N$ to the limit of an infinite system to obtain $P_{\nu,\gamma}^{(K)}(s,m)=\lim_{1/N\rightarrow0} P_{\nu,\gamma;N}^{(K)}(s,m)$.

The maximum feasible dimension of about $4\times 10^9$ meant that for simple fractions we have data for many system sizes (e.g., for $\nu=2/3$: $N\le28$ for the polarized phase and $N\le14$ for the unpolarized phase). However, for the more complex fractions such as $\nu=4/7$ and $4/9$ we only have data for two sizes, with the smaller size suffering from the ``aliasing'' problem (e.g., for $\nu=4/9$: $N=16$ and 20 for the polarized phase, $N=10$ and 14 for the partially polarized phase, and $N=8$ and 12 for the unpolarized phase). The results for these systems are therefore less reliable. The fractions $n/(2n\pm 1)$ with $n\geq 5$ are not amenable to exact diagonalization studies.

The extrapolated amplitudes were then convoluted with the effective pair and triplet LL mixing pseudopotentials $V_{\rm corr}^{(K)}(s,m)$ derived in Ref.~\onlinecite{Peterson13} (see Tables III and IV of that article) to give LL mixing corrections $\Delta\varepsilon_{\nu,\gamma}$ in the ``linear'' regime. In this regime, the effect of LL mixing is estimated perturbatively, to the first order in $\lambda$. Our estimates of LL mixing corrections to the ground state energy per particle are given per unit of $\lambda$, separately for each filling factor and spin polarizarion: $\Delta\varepsilon_{\nu,\gamma}/\lambda=\sum_{K;s,m}P_{\nu,\gamma}^{(K)}(s,m) V_{\rm LL\, mix}^{(K)}(s,m)$, with the sums running over all spins and over the leading angular momenta for which the pseudopotentials are available\cite{Peterson13} (however, since $m$ corresponds to an average $K=2$ body distance or $K=3$ body area, both $K=3$ sums (for $s=1/2$ and $3/2$) ought to be limited to the same $m_{\text{max}} = 3$.).

(It should also be mentioned that alternative methods for including LL mixing are in principle possible. For example, one can attempt diagonalization of the $N$-electron Hamiltonian in an expanded Hilbert space, including cyclotron-excited configurations with some occupation of higher Landu levels.\cite{Wojs06,Rezayi11} However, we have not found this method to be feasible for the present problem.)

\begin{figure*}
\begin{center}
\includegraphics[width=0.8\textwidth,height=0.45\textwidth]{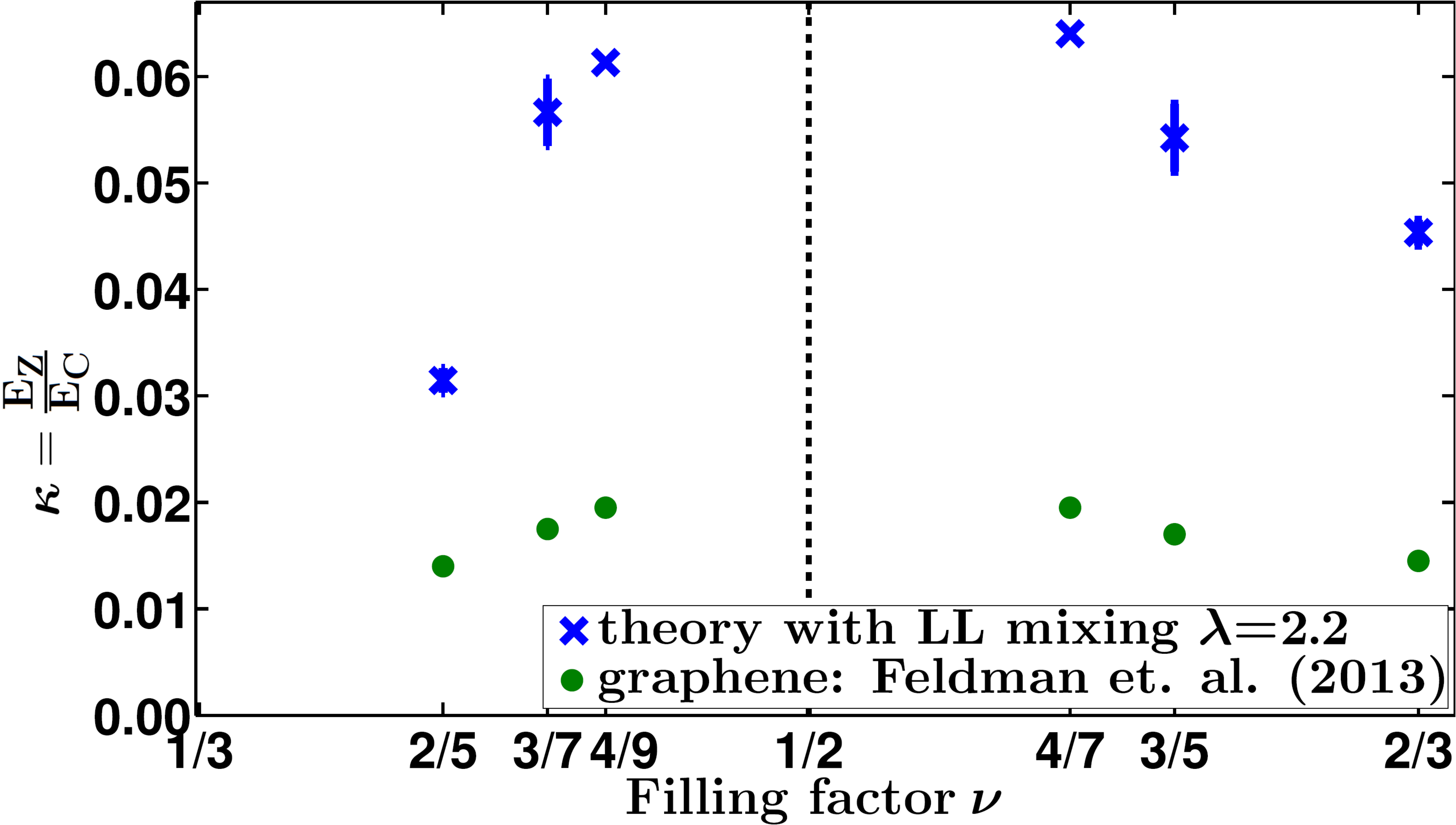}
\caption{(Color online) This figure compares the experimental phase diagram in graphene (taken from Feldman {\em et al.}\cite{Feldman13}, shown by green dots) with the theoretical phase diagram (blue crosses) including corrections from LL mixing, assuming that the correction remains linear in $\lambda$. The data of Feldman {\rm et al.} \cite{Feldman13} assumes $g\epsilon=6.0$. The theoretical estimates were obtained with $\lambda=2.2$, which corresponds to the suspended graphene samples of Feldman {\rm et al.} \cite{Feldman13}. (We note that the blue crosses for the spin transitions at $\nu=4/7$ and $\nu=4/9$ are obtained using results extrapolated from only two finite systems.) }
\label{fig:LLmix}
\end{center}
\end{figure*}

The modified critical values of $\kappa$ for $\lambda=1$ are shown in the tables in the Appendix \ref{appendix1} for $\nu=2/3$, 3/5, 4/7, 2/5, 3/7 and 4/9. A comparison with the experimental values is shown in Fig.~\ref{fig:LLmix} where in the experimental data was obtained by assuming $g\epsilon=6$. A better agreement between the theoretical and experimental results can be obtained by choosing the value of $g\epsilon=16$, but this value seems implausible. Therefore we come to the conclusion that theory substantially overestimates the effect of LL mixing. There may be several possible origins for this. 

First, the correction to the interaction has been calculated to linear order in $\lambda$ perturbatively, and is thus valid only so long as the correction to the energy remains linear in $\lambda$. It is possible that $\lambda = 2.2$ is outside the linear regime. We have also assumed that the wave functions themselves are not significantly modified by LL mixing. This should be the case for small $\lambda$ but may not be valid for $\lambda=2.2$. Finally, we have included corrections only for pseudopotentials up to a given relative angular momentum. The quantitative errors due to such an {\em ad hoc} truncation are not known but may be significant. We believe that these comparisons bring out complications associated with the theoretical treatment of the quantitative effect of LL mixing. 

\section{Composite fermions carrying four vortices}

It was predicted in Ref.~\onlinecite{Park99} that the spin physics of the FQHE states at $n/(4n+1)$, described in terms of composite fermions carrying four vortices ($^4$CFs), is qualitatively different from that at $n/(2n\pm 1)$. Calculations based on the JK wave functions indicated that the fully spin polarized state $(n,0)$ at $n/(4n+1)$ is the ground state even at zero Zeeman energy, and consequently there are no spin-polarization phase transitions. The failure of the free-CF model was interpreted in Ref.~\onlinecite{Park99} in terms of a Bloch ferromagnetism for composite fermions, caused by a large exchange interaction that dominates their CF-cyclotron energy and favors the fully polarized state even in the absence of the Zeeman energy.

We have seen above that the JK projection overestimates the energies of the non-fully spin polarized state by a larger amount than of the fully spin polarized states. One may therefore ask if the result in Ref.~\onlinecite{Park99} is an artifact of the JK projection scheme. Furthermore, the spin physics at $n/(4n-1)$ has not been investigated so far. Are these states also always fully spin polarized? 

\begin{figure*}[htpb]
\begin{center}
\includegraphics[width=0.66\columnwidth,height=0.36\columnwidth]{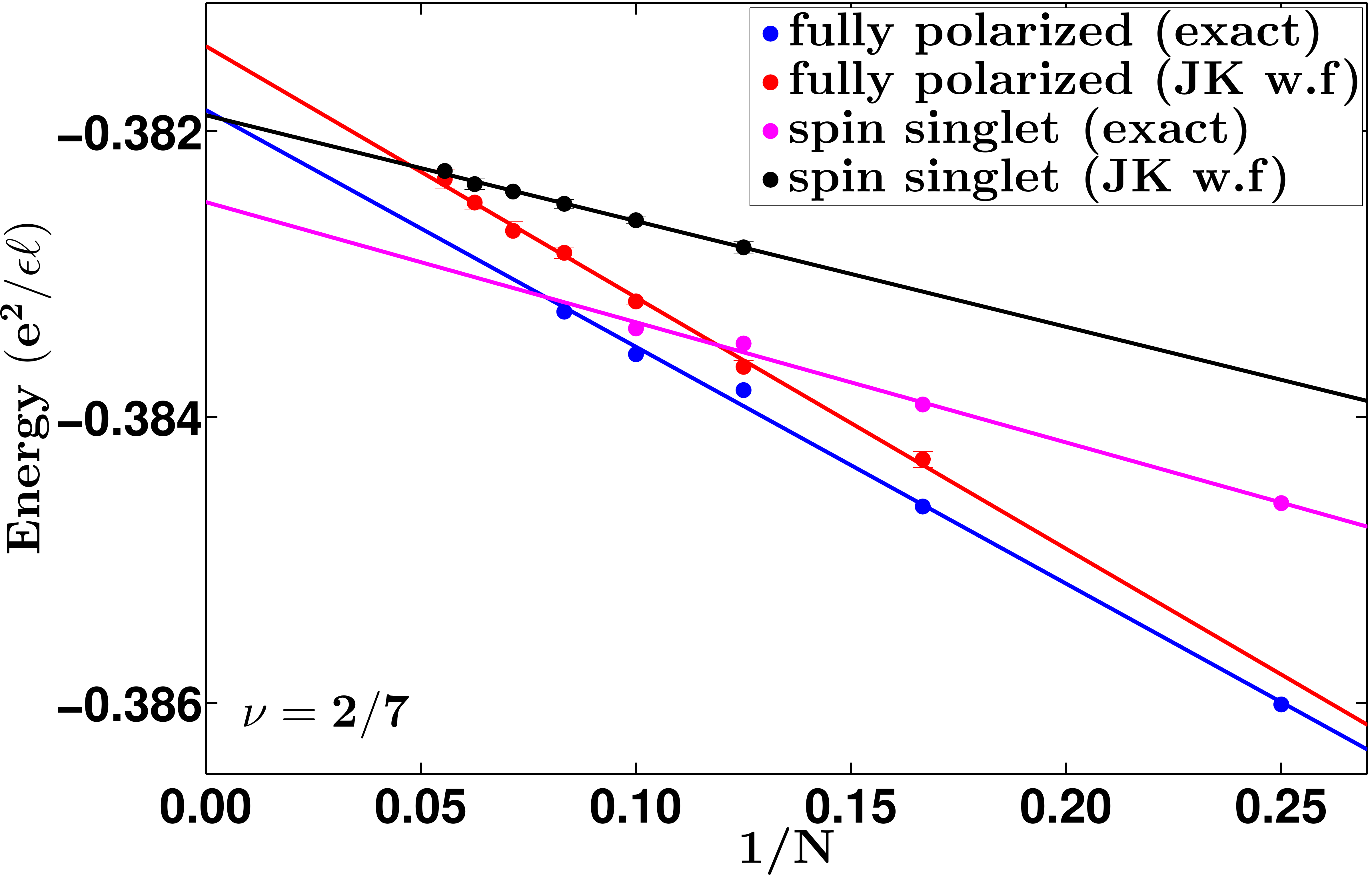}
\includegraphics[width=0.66\columnwidth,height=0.36\columnwidth]{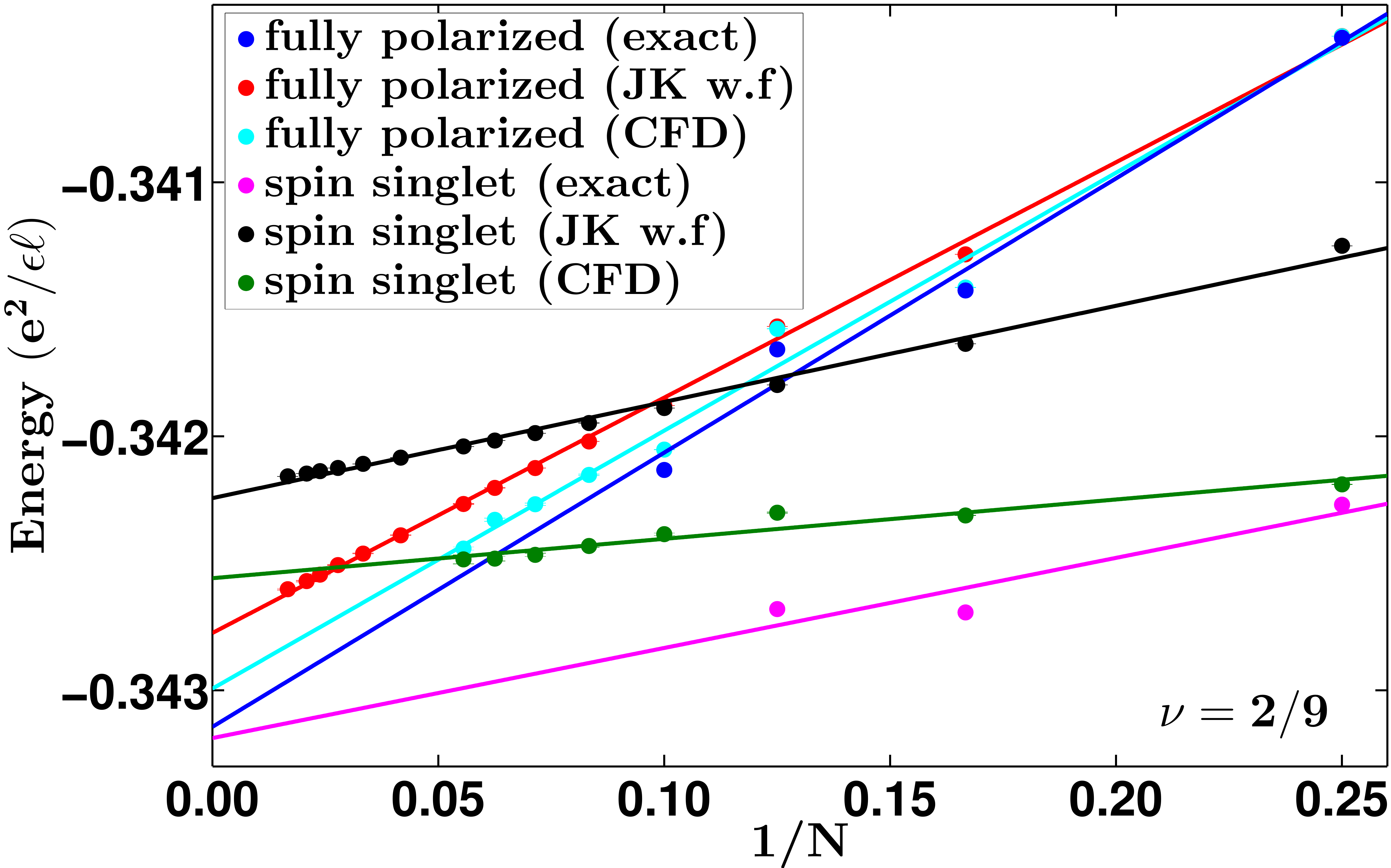}

\includegraphics[width=0.66\columnwidth,height=0.36\columnwidth]{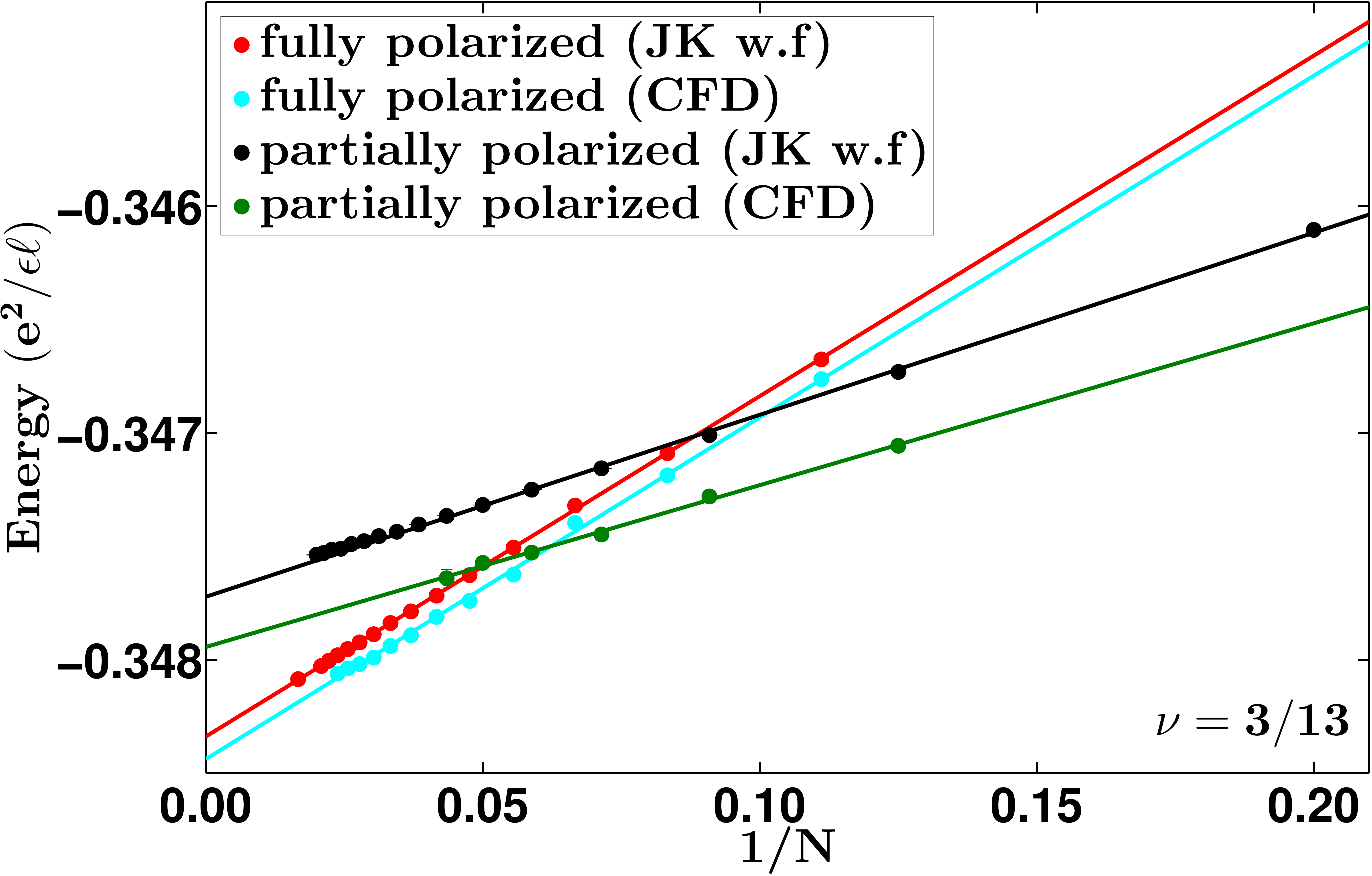}
\includegraphics[width=0.66\columnwidth,height=0.36\columnwidth]{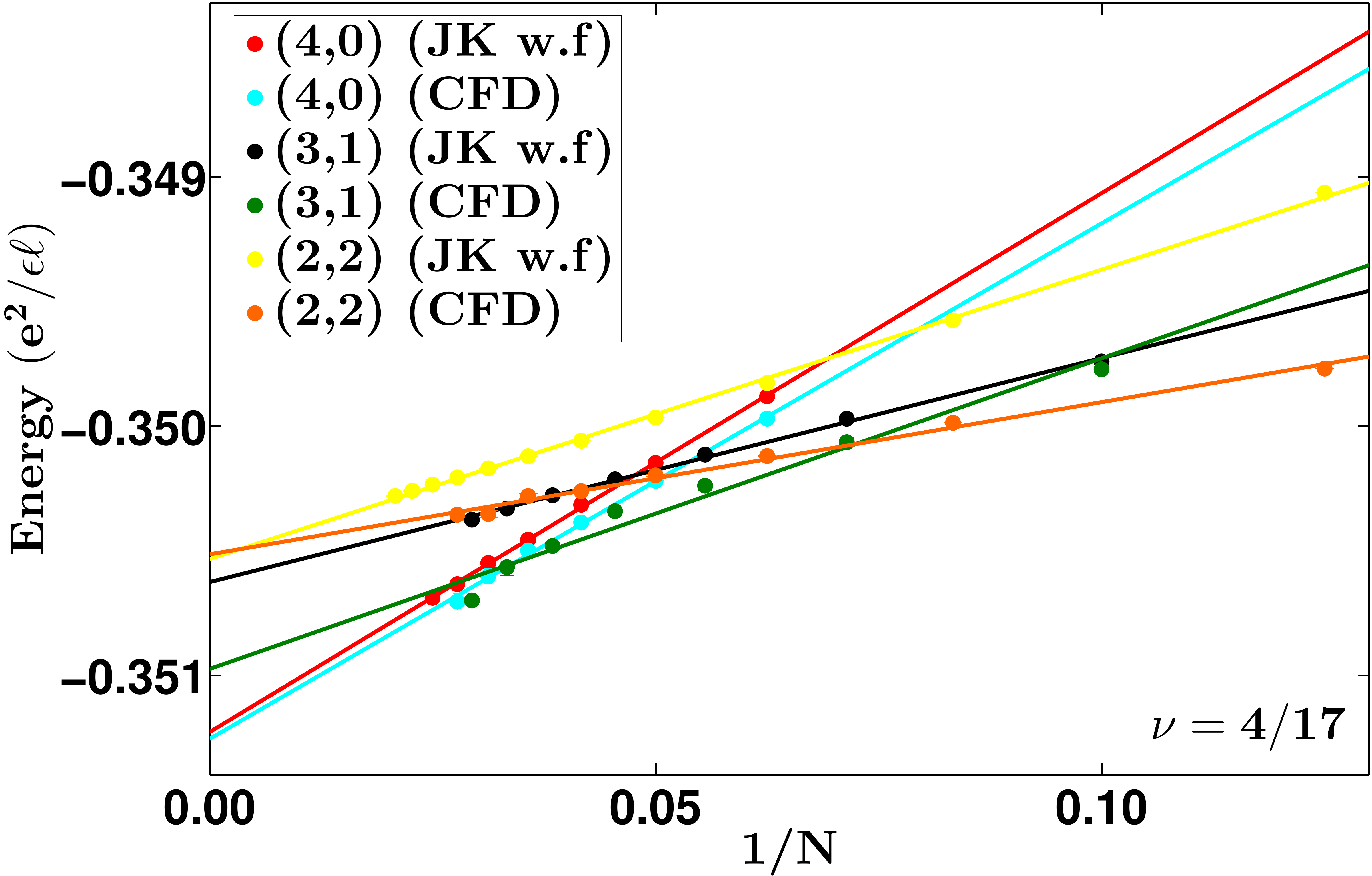}
\caption{(Color online) Thermodynamic extrapolation of the Coulomb ground state energies at $\nu=2/7$ (top left panel), $2/9$ (top right panel), $3/13$ (bottom left panel) and $4/17$ (bottom right panel). The energies are obtained from exact diagonalization (``exact"), JK wave functions (``JK w.f.") or CF diagonalization (``CFD").}
\label{fig:4CFs}
\end{center}
\end{figure*}

We have investigated these questions both by exact and CF diagonalizations for some of these fractions. For 2/9 FQHE state, exact diagonalization results are inconlusive but both the JK wave function and CF diagonalization (see Fig. \ref{fig:4CFs}) find that the fully spin polarized state has lower energy, thus confirming absence of any spin-polarization phase transition. The same is true for the $3/13$ and $4/17$ FQHE states wherein both the JK wave function and CF diagonalization find the state with larger spin polarization to be lower in energy than the state with a smaller spin polarization (see Fig. \ref{fig:4CFs}). We have also studied the first meaningful member of the Jain sequence $n/(4n-1)$, namely 2/7. Here, the calculations based on the JK wave functions as well as exact diagonalization show that the spin singlet state has a slightly lower energy (see Fig. \ref{fig:4CFs}), giving $\kappa_{2/7}=0.0013(3)$ (exact).

\begin{figure}[htbp]
\begin{center}
\includegraphics[width=0.49\columnwidth]{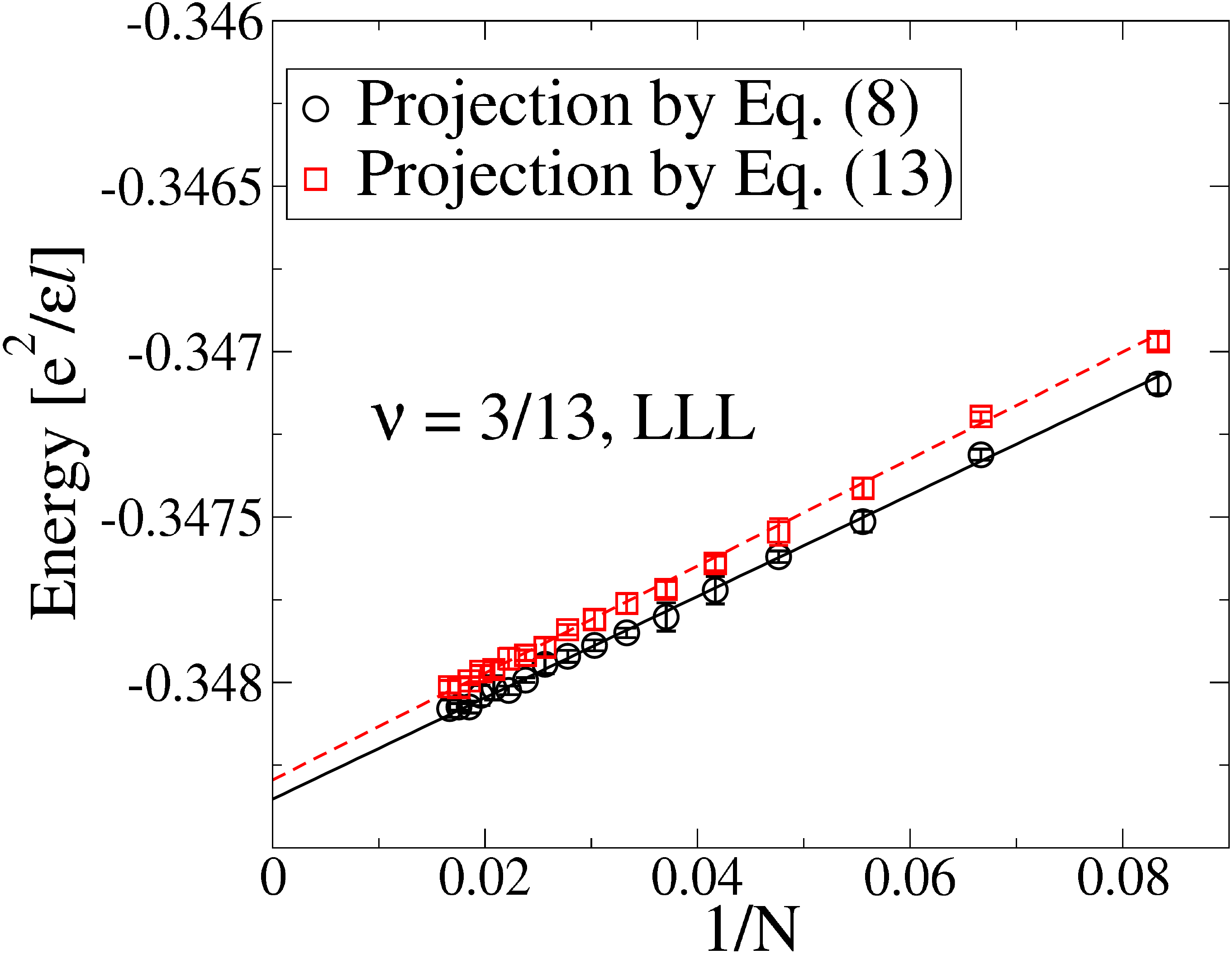}
\includegraphics[width=0.49\columnwidth]{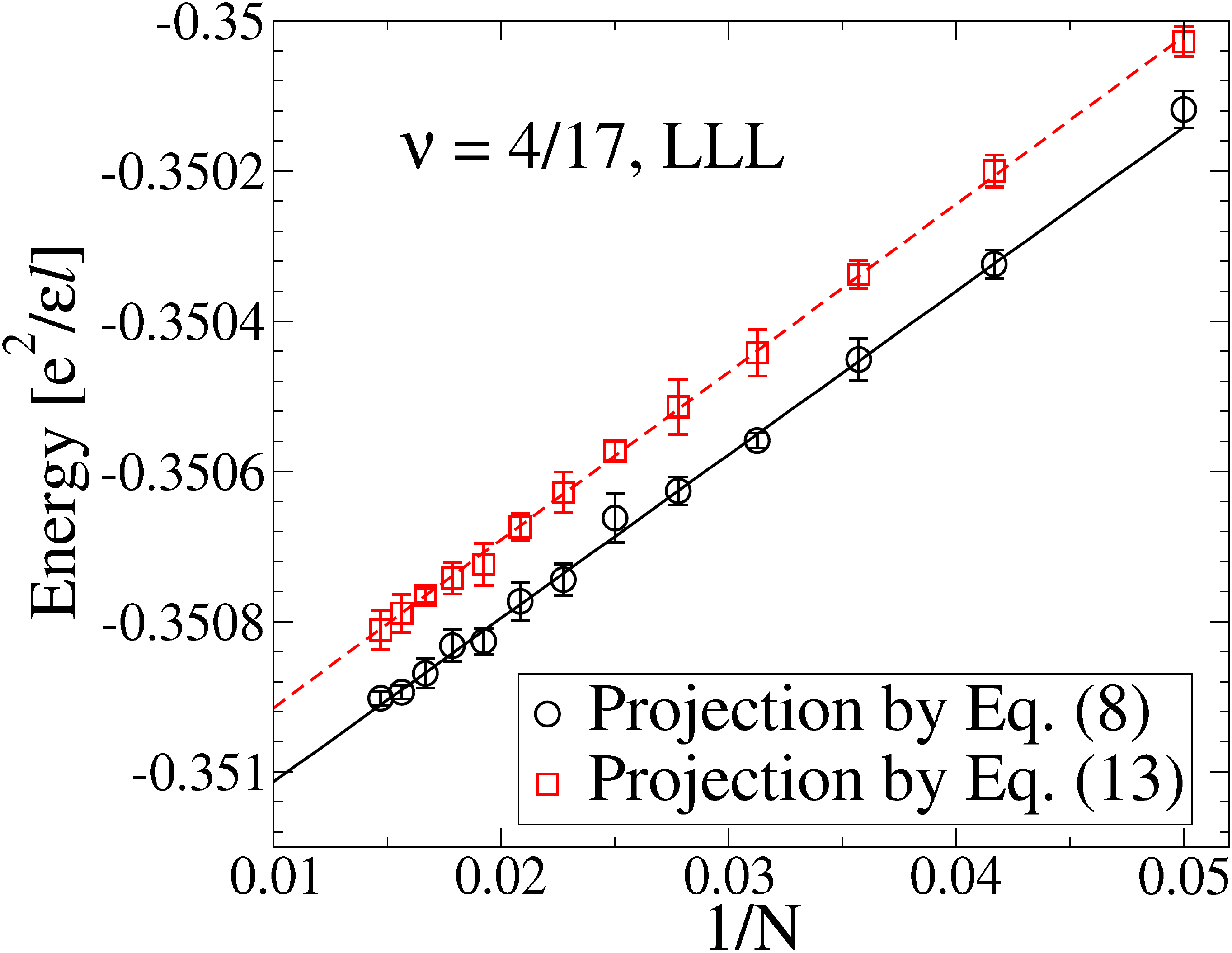}
\end{center}
\caption{\label{fig:paraproj}
Comparison of the LLL Coulomb ground state energies obtained from wave functions of Eq.~\ref{paral-flux} (black squares) and Eq.~\ref{diff-proj_paral} (red squares) for fully polarized states at $3/13$ (left panel) and $4/17$ (right panel). The former gives better energies and is used for the results in Tables \ref{tab:n_2} to \ref{tab:n_6}.}
\end{figure}

\begin{figure}[htbp]
\begin{center}
\includegraphics[width=0.49\columnwidth]{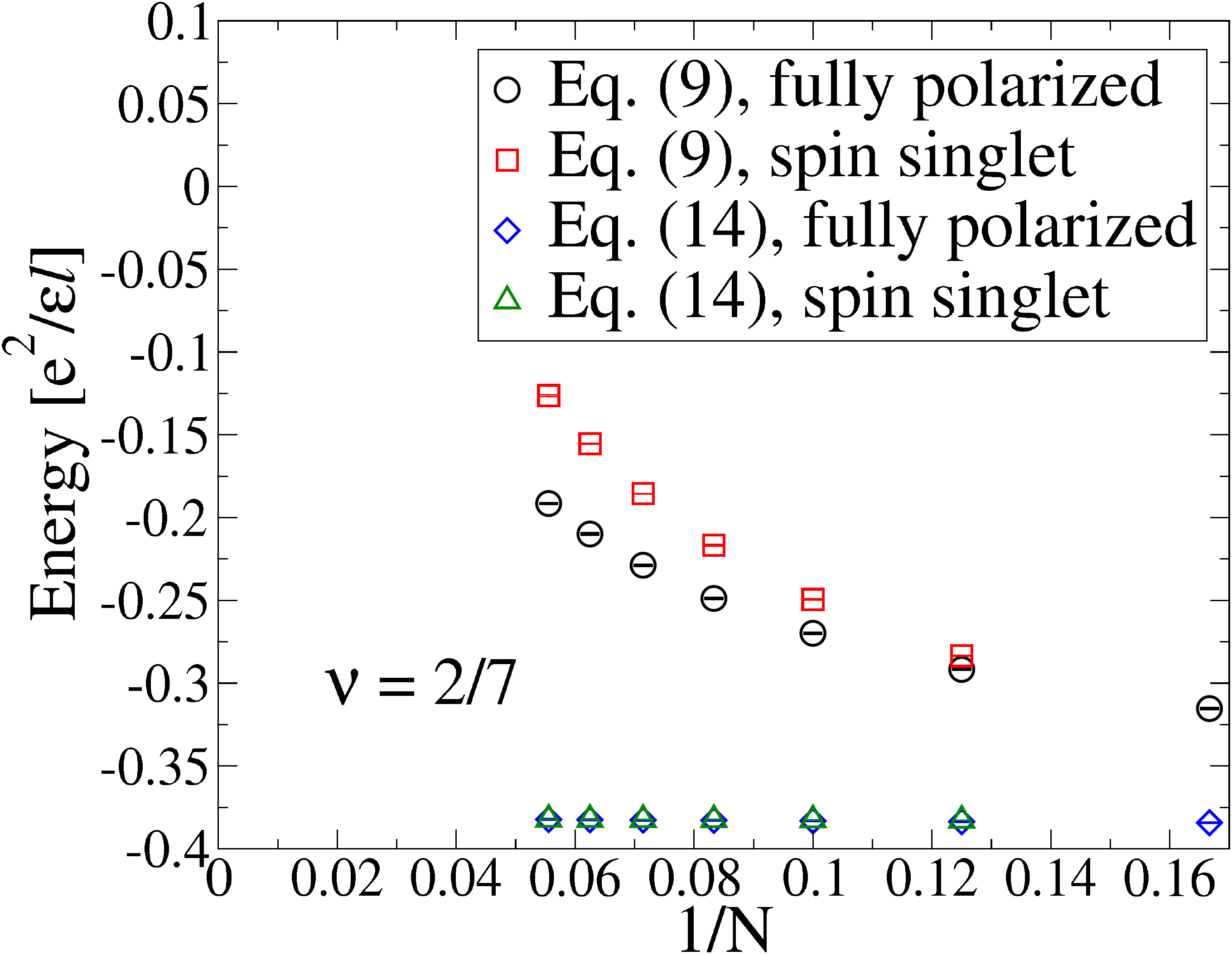}
\includegraphics[width=0.49\columnwidth]{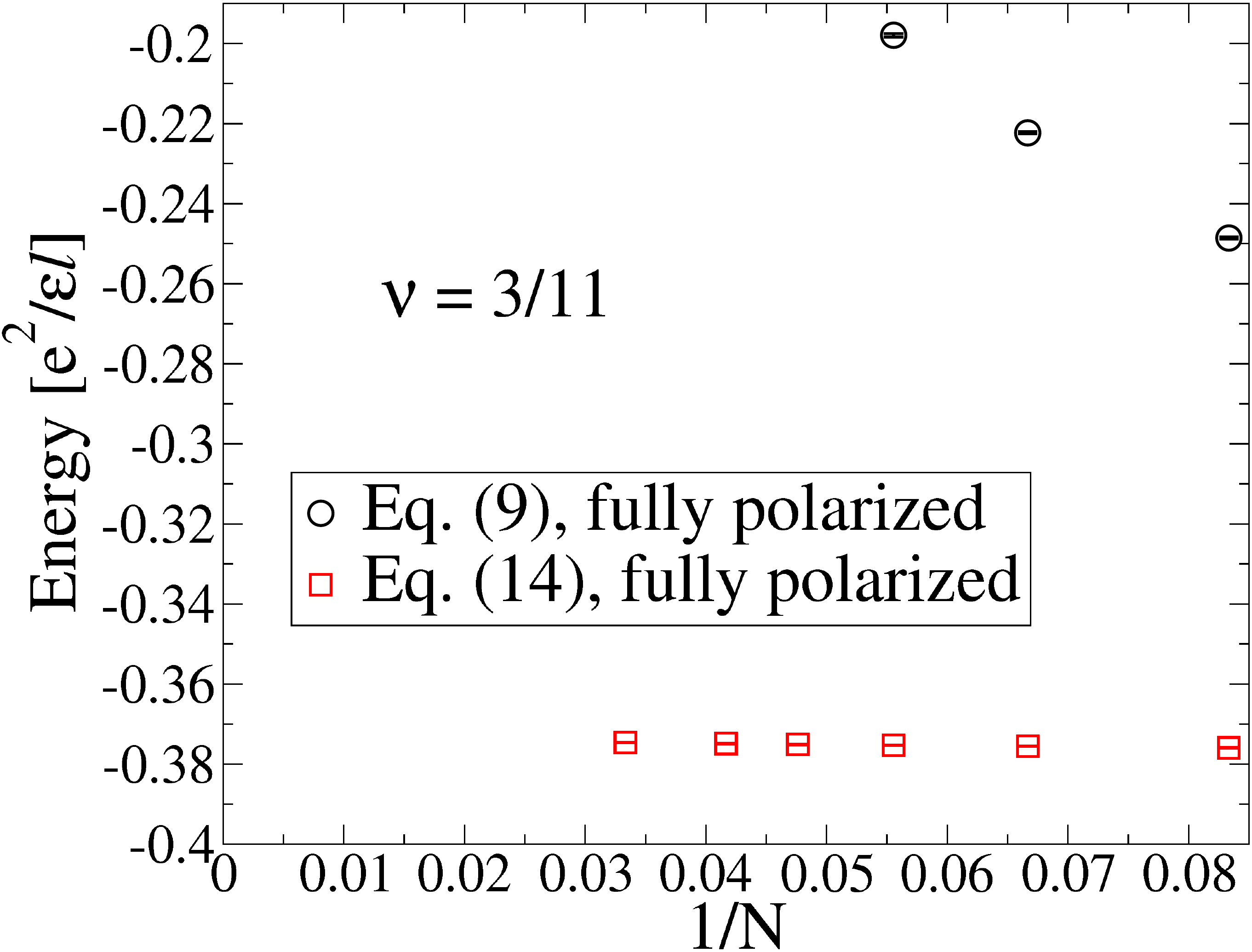}
\end{center}
\caption{\label{fig:revproj}
Comparison of the LLL Coulomb ground state energies obtained from wave functions of Eq.~\ref{rev-flux} and Eq.~\ref{diff-proj_rev} for states with reverse flux attachment. The latter gives better energies, and is used to obtain the energies shown in Tables \ref{tab:n_2} to \ref{tab:n_6}.}
\end{figure}

For the $^{2p}$CF states with $p>1$, one can also perform the JK projection slightly differently as:
\begin{eqnarray}
\Psi_{\frac{n}{2pn+1}}=J^{2p-2} {\cal P}_{\rm LLL} \Phi_{n} J^{2} 
\label{diff-proj_paral}\\
\Psi_{\frac{n}{2pn-1}}=J^{2p-2} {\cal P}_{\rm LLL} [\Phi_{n}]^{*} J^{2} 
\label{diff-proj_rev}
\end{eqnarray}
Because these wave functions apparently build better short distance correlations, one may expect them to have lower energies than their counterparts in Eqs.~(\ref{paral-flux}) and (\ref{rev-flux}). Contrary to this expectation we find that at $\nu=n/(2pn+1)$ the wave functions in Eq.~\ref{diff-proj_paral} in general have slightly higher energies for finite systems than those in Eqs.~\ref{paral-flux}, as seen in Fig.~\ref{fig:paraproj}. (We note that for states restricted to two $\Lambda$ levels, the wave functions of Eq.~\ref{diff-proj_paral} and Eq.~\ref{paral-flux} are identical.). In contrast, for states at $\nu=n/(2pn-1)$ we find that the ground state energies obtained from the wave functions in Eq.~\ref{diff-proj_rev} are lower than those obtained from Eq.~\ref{rev-flux} (see Fig.~\ref{fig:revproj}). Therefore, for states at $\nu=n/(2pn+1)$ with $p>1$ we quote energies obtained from Eqs.~(\ref{paral-flux}) while for states at $\nu=n/(2pn-1)$ we quote energies obtained from Eq.~\ref{diff-proj_rev}.

\section{Conclusions} 

The critical Zeeman energies where transitions between differently spin polarized states occur are a direct measure of the energy difference between the states, and thus serve as a very sensitive test of the quantitative accuracy of the theory of the FQHE. These critical Zeeman energies are in general subject to corrections due to finite thickness of the quantum well and also Landau level mixing.  In GaAs both effects are present, and it is not straightforward to disentangle their contributions, although progress has been made in experiments that study the effect systematically as a function of the quantum well width \cite{Liu14}.  Because graphene has negligible finite thickness corrections, this gives an opportunity to obtain an accurate test of the CF theory, and also to gain insight into our understanding of the role of LL mixing \cite{Peterson14}.

We have evaluated accurate spin-polarization phase diagram for the FQHE for an ideal two-dimensional system confined to the LLL with no LL mixing and no disorder. We have also evaluated corrections due to LL mixing, assuming that these are linear in the parameter $\lambda$. We find that the experimental results of Feldman {\em et al.} \cite{Feldman13} are in excellent agreement with theory that neglects LL mixing. Somewhat unexpectedly, if we include LL mixing in a linear approximation, the agreement becomes significantly worse, indicating that the amount of LL mixing in experiments is too large to be captured by a first order perturbative treatment. These results underscore our lack of a quantitative understanding of the effect of LL mixing on various quantities. 

We have shown that the critical Zeeman energies are well captured in terms of an effective mass model of composite fermions. We have shown that the CF states with Jain-Kamilla projection produce the correct energy ordering for these states, and are fairly accurate for the ``parallel flux attached" states at $\nu=n/(2n+1)$, producing the critical Zeeman energies with better than $\sim$15\% accuracy. In contrast, for the ``reverse flux attached" states at $\nu=n/(2n-1)$ the JK projection obtains the correct energy ordering of the states but obtains the critical Zeeman energies that are off by a factor of two to three. For these states, the hard-core projection method is very accurate but difficult to implement for large systems.

\section{Acknowledgment}
The work was supported by the U.S. Department of Energy, Office of Science, Basic Energy Sciences, under Award No. DE-SC0005042 (ACB, JKJ),  
Hungarian Scientific Research Funds No. K105149 (CT), the Polish NCN grant 2014/14/A/ST3/00654 and the EU Marie Curie Grant PCIG09-GA-2011-294186 (AW). We thank Research Computing and Cyberinfrastructure at Pennsylvania State University (supported in part through instrumentation funded by the National Science Foundation through grant OCI-0821527), the HPC facility at the Budapest University of Technology and Economics and Wroc{\l}aw Centre for Networking and Supercomputing and Academic Computer Centre CYFRONET, both parts of PL-Grid Infrastructure. Csaba T\H oke was supported by the Hungarian Academy of Sciences. 

\appendix
\section{Results}
\label{appendix1}
\begin{figure*}[htbp]
\begin{center}
\includegraphics[width=0.66\columnwidth,height=0.38\columnwidth]{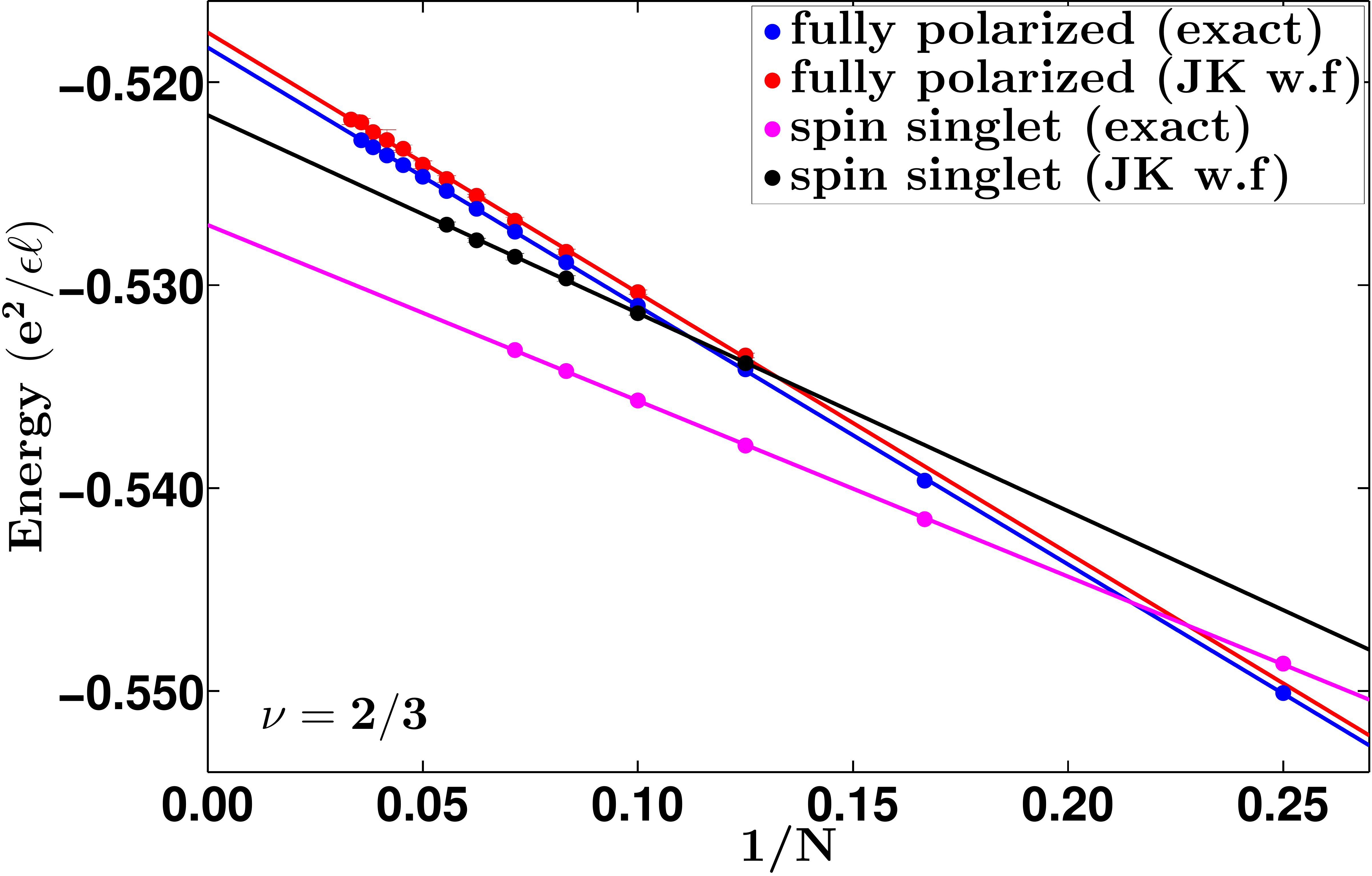}
\includegraphics[width=0.66\columnwidth,height=0.38\columnwidth]{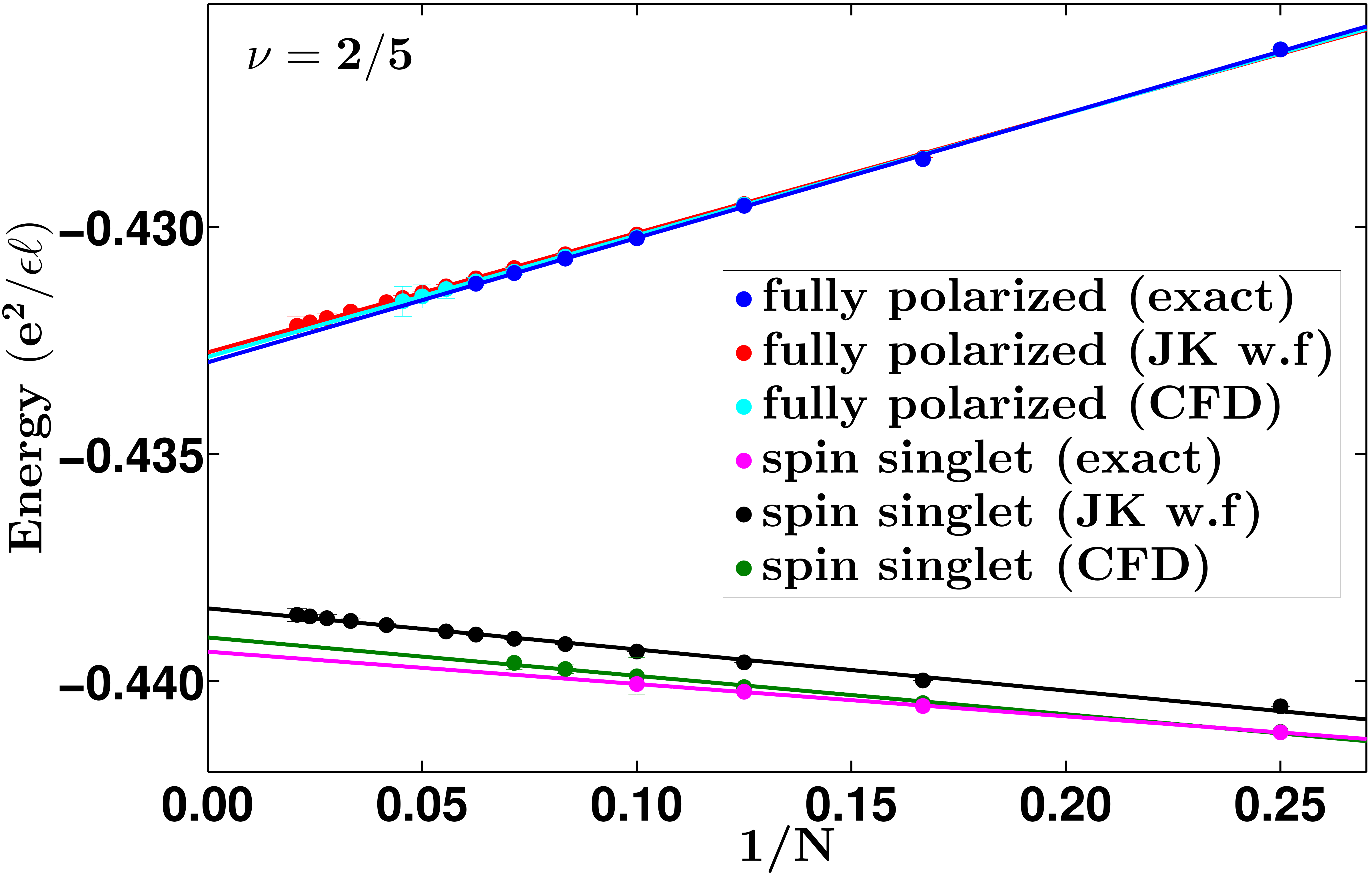}
\includegraphics[width=0.66\columnwidth,height=0.38\columnwidth]{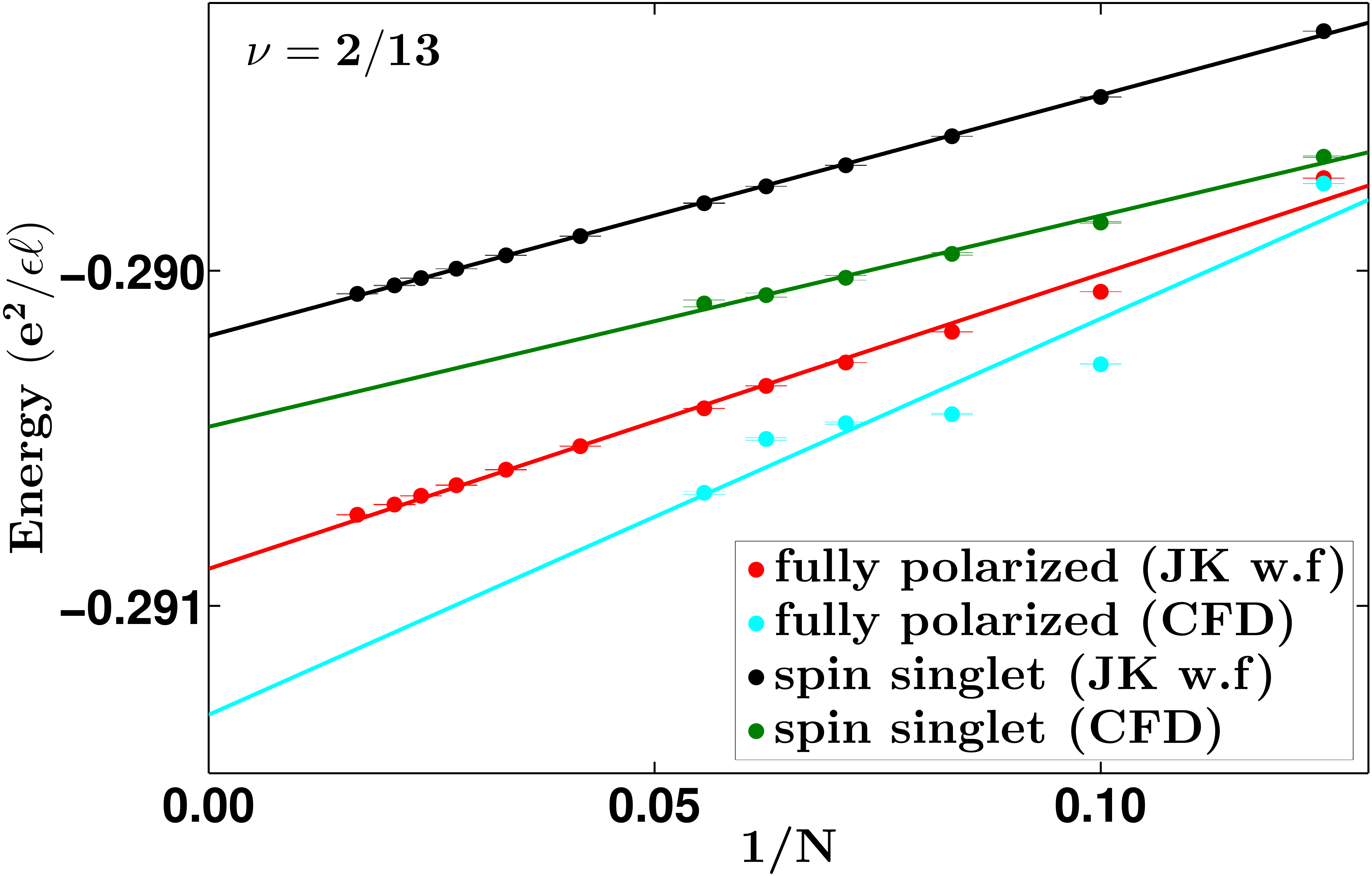}

\includegraphics[width=0.66\columnwidth,height=0.38\columnwidth]{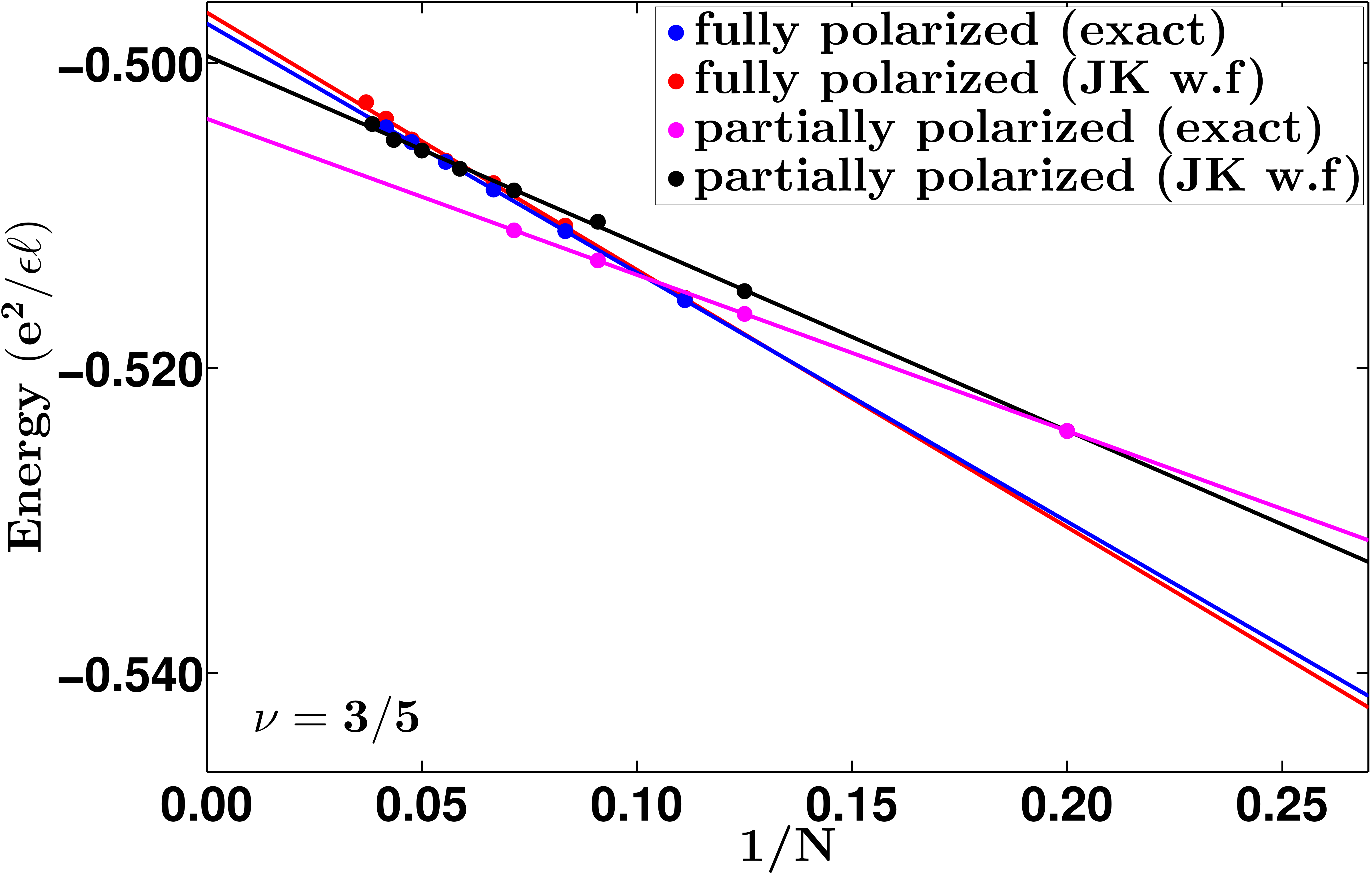}
\includegraphics[width=0.66\columnwidth,height=0.38\columnwidth]{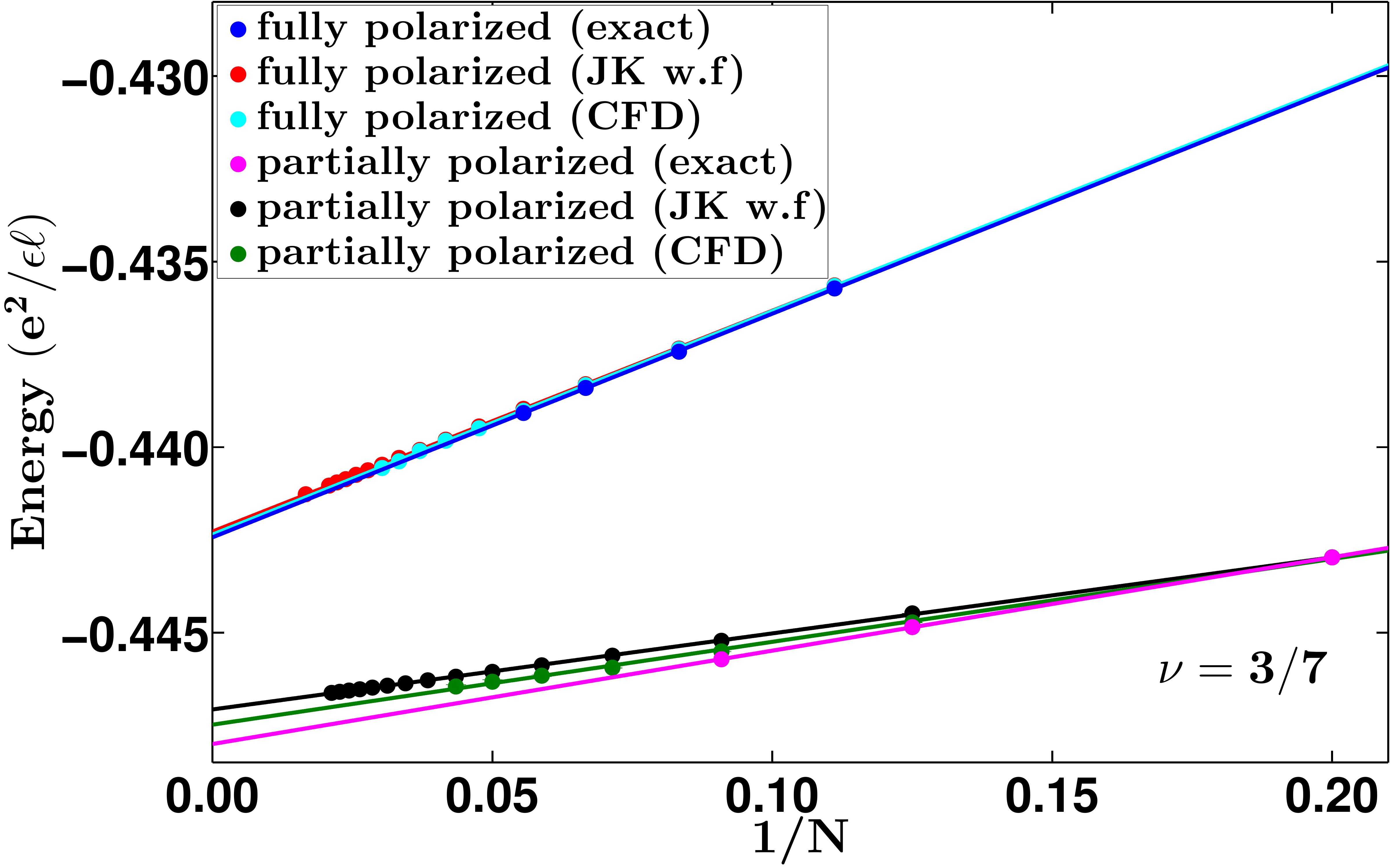}  
\includegraphics[width=0.66\columnwidth,height=0.38\columnwidth]{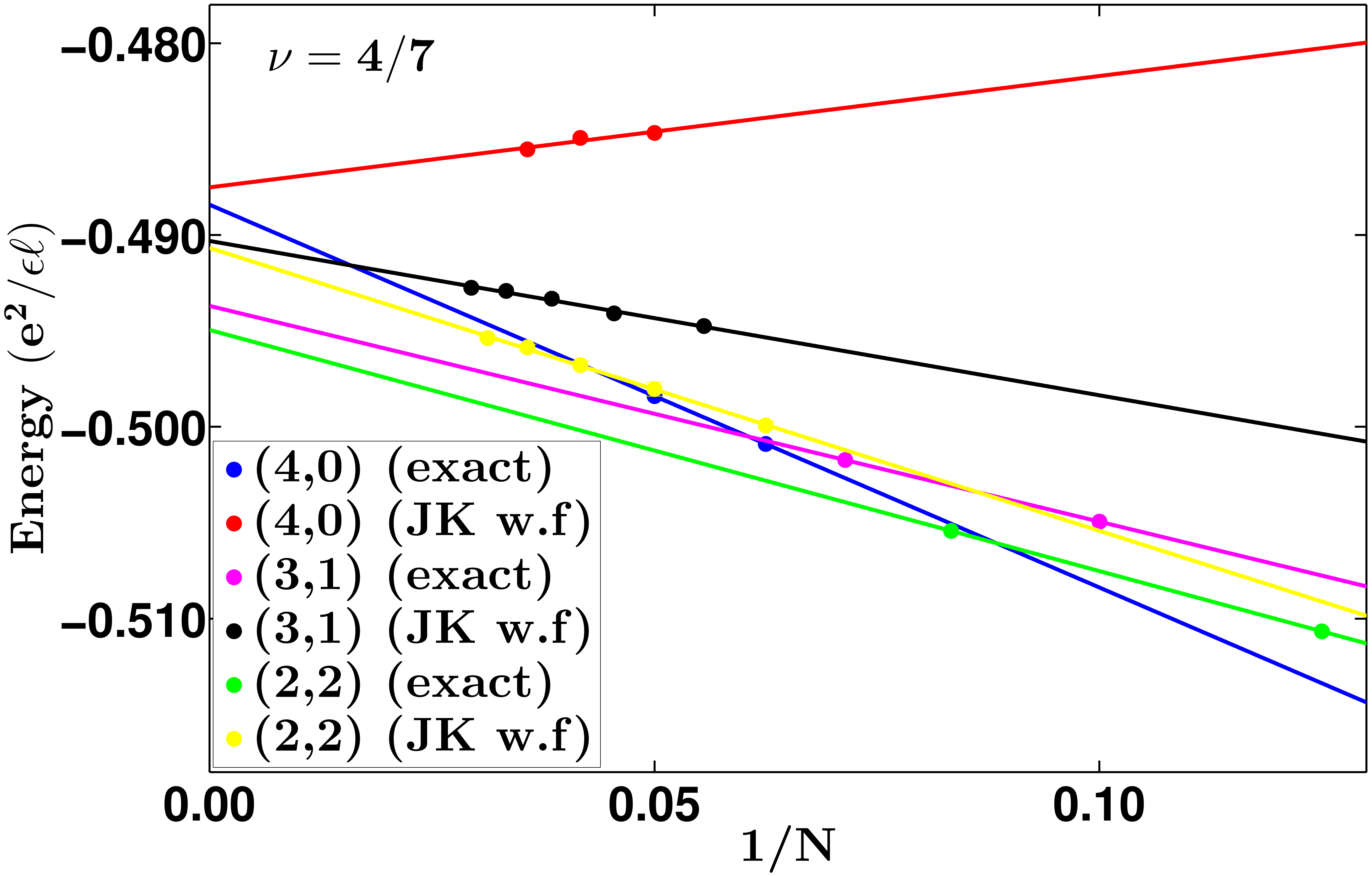}  

\includegraphics[width=0.66\columnwidth,height=0.38\columnwidth]{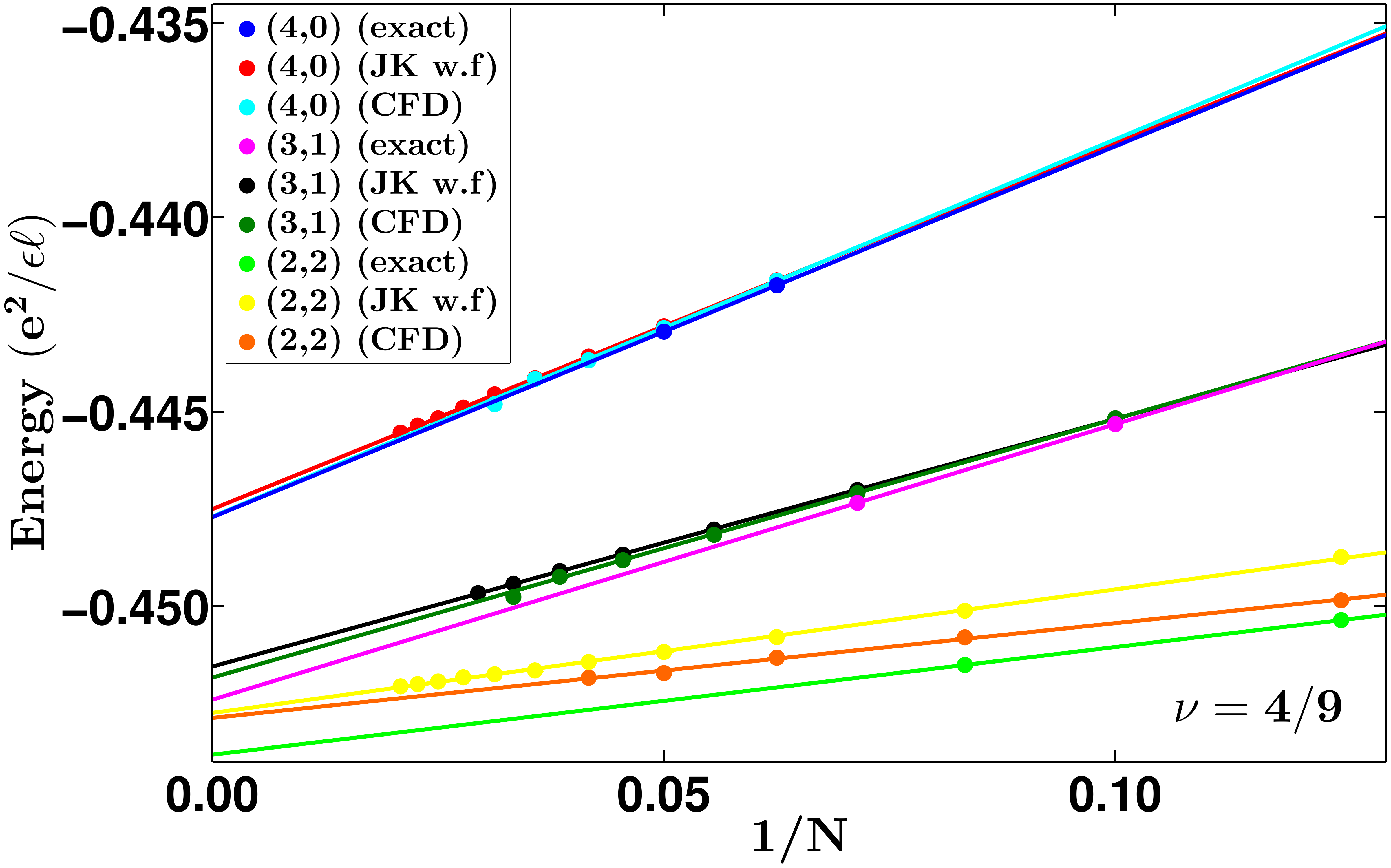}
\includegraphics[width=0.66\columnwidth,height=0.38\columnwidth]{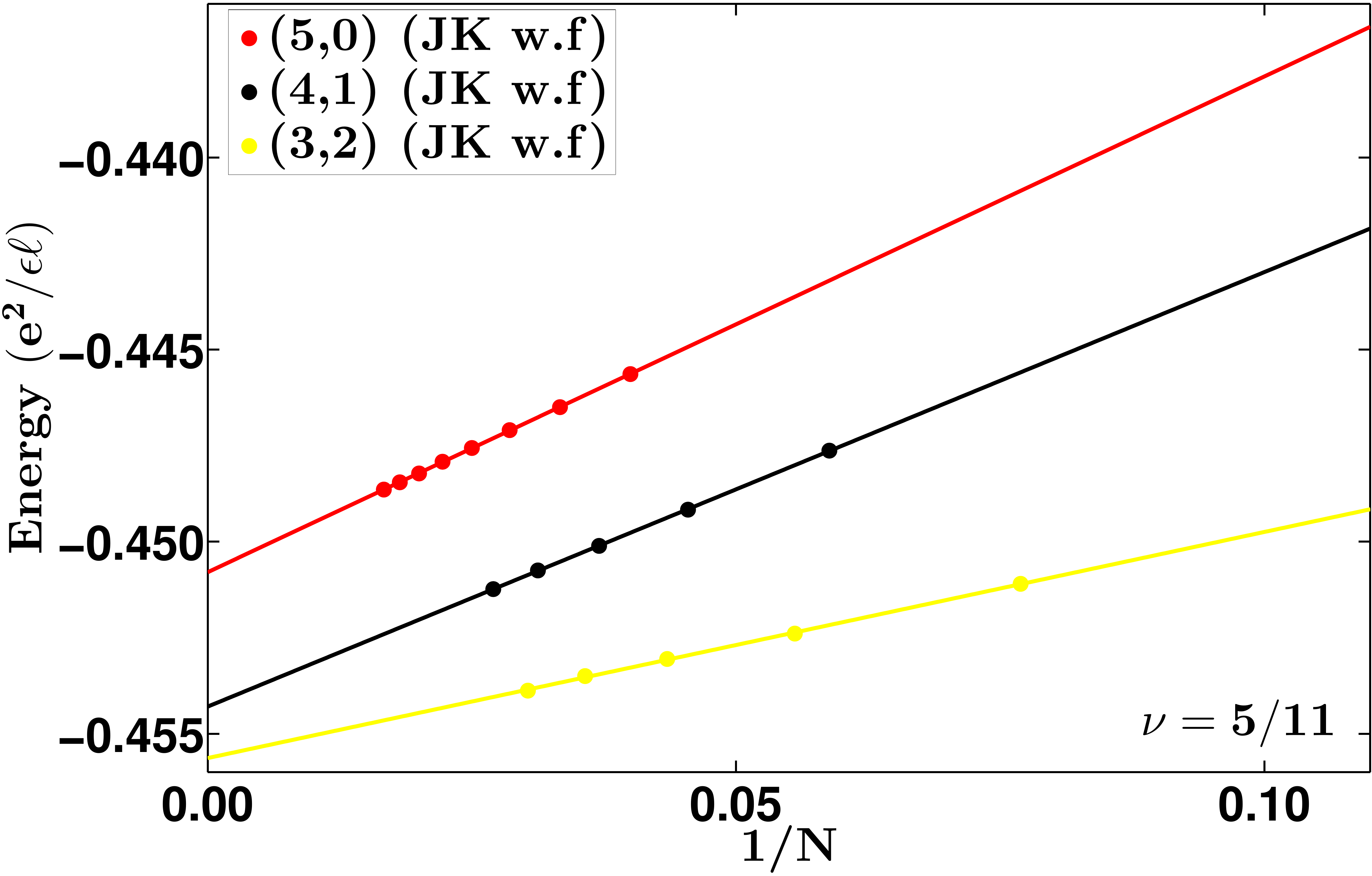}
\includegraphics[width=0.66\columnwidth,height=0.38\columnwidth]{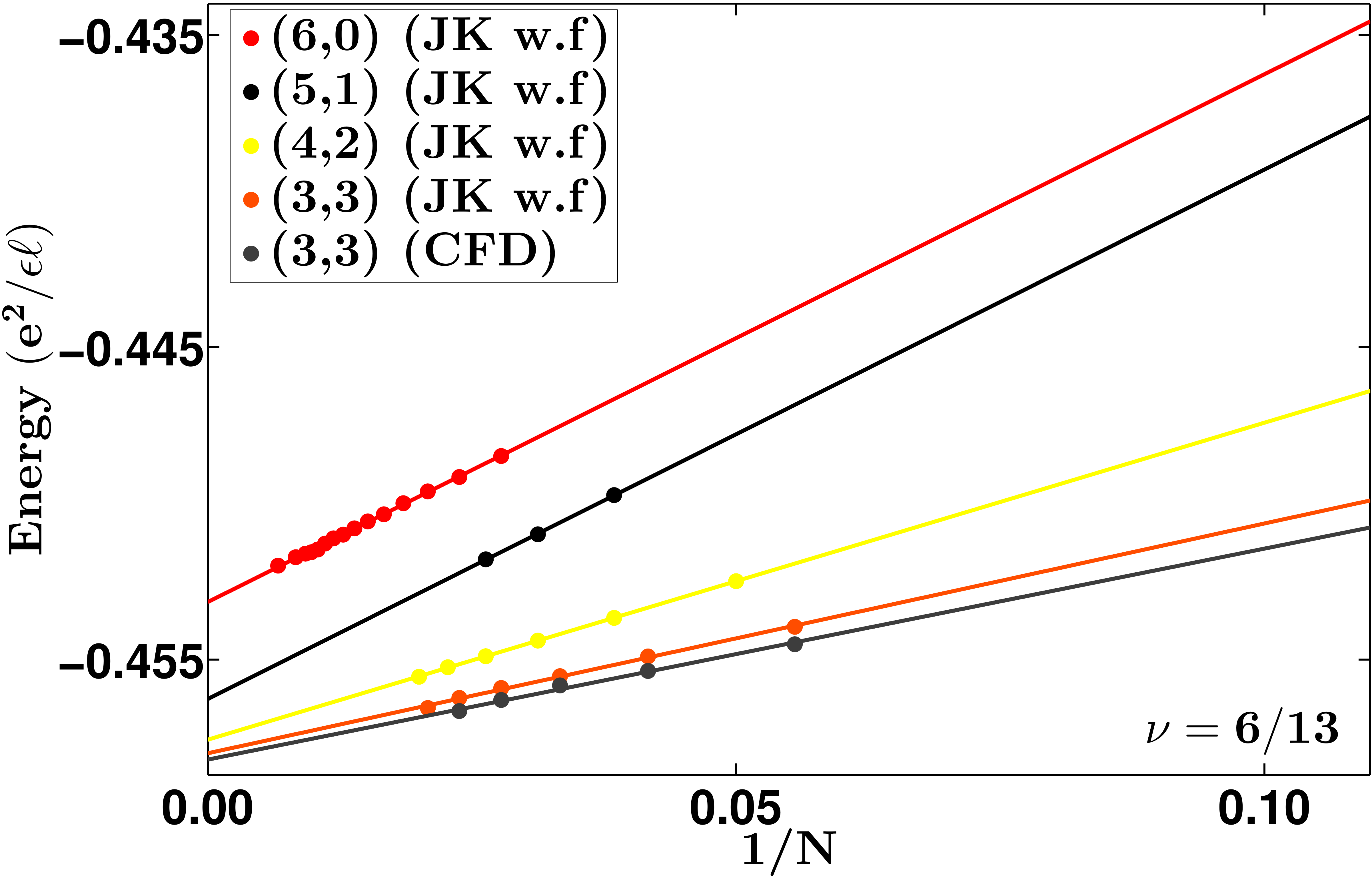}
\end{center}
\caption{\label{fig:extrapol}
Extrapolation of the lowest Landau level Coulomb ground state energy to the thermodynamic limit for different spin polarized states at various filling factors.}
\end{figure*}

In this Appendix, we give the results for the individual systems used in obtaining the extrapolated energies shown in the main text. The tables below give energies obtained from three methods. One is exact diagonalization in the spherical geometry. These results are obtained by the Lanczos method. In Table \ref{tab:exact} we give the dimensions of the full lowest Landau level Hilbert space for the largest system size at various filling factors considered in this work. These are the dimensions for the $L=0$ and the relevant $S$ sector. The results labeled ``JK w.f" are obtained from Jain's CF wave functions using the JK projection method while those labeled ``CFD" are obtained by the method of composite fermion diagonalization \cite{Mandal02}. In Fig. \ref{fig:extrapol} we show the thermodynamic extrapolation of the ground state energies from finite size calculations of different spin polarized states at various filling factors in the lowest Landau level. Tables \ref{tab:n_2}, \ref{tab:n_3}, \ref{tab:n_4}, \ref{tab:n_4_EZ}, \ref{tab:n_5} and \ref{tab:n_6} show the thermodynamic energies obtained from these extrapolations and the critical Zeeman energies for the spin transitions. In these tables we also show the corrections obtained from LL mixing for the LL mixing parameter value $\lambda=1$.

\begin{table}
\begin{tabular}{|c|c|c|c|c|c|c|}
\hline
\multicolumn{1}{|c|}{~~$\nu$~~} & \multicolumn{1}{|c|}{~~state~~} & \multicolumn{1}{|c|}{~~$N$~~} & \multicolumn{1}{|c|}{~~$2Q$~~} & \multicolumn{1}{|c|}{~~$S$~~} & \multicolumn{1}{|c|}{~~dimension~~} \\ \hline
2/3	& (2,0)   	& 28 & 42 &  14	& 1,521,967,986	\\ \hline 
2/3	& (1,1)		& 14 & 20 &  0	&   280,934,870	\\ \hline 

2/5	& (2,0)		& 18 & 41 &  9	& 3,546,374,322	\\ \hline 
2/5	& (1,1)		& 12 & 27 &  0	& 2,211,680,688	\\ \hline  

3/5	& (3,0)		& 24 & 41 &  12	& 3,546,374,322	\\ \hline 
3/5	& (2,1)		& 14 & 23 &  3	&   383,215,178	\\ \hline    

3/7	& (3,0)		& 18 & 37 &  9	&   386,905,330	\\ \hline 
3/7	& (2,1)		& 11 & 22 &  2.5	&    17,969,272	\\ \hline 

4/7	& (4,0)		& 20 & 37 &  10	&   386,905,330	\\ \hline 
4/7	& (3,1) 	& 14 & 25 &  5	&    55,975,102	\\ \hline 
4/7	& (2,2)		& 12 & 21 &  0	&   114,153,021	\\ \hline

4/9	& (4,0)		& 20 & 39 &  10	& 1,438,058,853	\\ \hline 
4/9	& (3,1)		& 14 & 27 &  5	&   186,301,264	\\ \hline 
4/9	& (2,2)		& 12 & 23 &  0	&   336,012,314	\\ \hline

\end{tabular}
\caption {Dimension of the Hilbert space of the largest systems for which exact diagonalization results were obtained in this work. We show the dimension of states in the $L=0$ and relevant $S$ sector for several values of $(N,2Q)$ at various filling factors.} 
\label{tab:exact} 
\end{table}

\begin{table*}
\begin{center}
\begin{tabular}{|c|c|c|c|c|c|c|c|c|c|}
\hline
\multicolumn{1}{|c|}{$\nu$} & \multicolumn{3}{|c|}{(2,0)} & \multicolumn{3}{|c|}{(1,1)} 	& \multicolumn{3}{|c|}{$\kappa$}  \\ \hline
	    & exact  		& JK w.f. 		&  CFD 		& 	exact	 & 	JK w.f. 	& CFD	 	& 	exact	& 	JK w.f. 	&  CFD		\\ \hline

2/3 	    & -0.51829(2) 	& -0.5176(1)	&		& -0.52704(4) & -0.5217(2)   	&-		&      0.0175(1)& 0.0065(17)	& -		\\ \hline
2/3 (with LL mixing)	    & -0.61469(16) 	& -		&		& -0.62969(10) & -	   	&-		&      0.0300(5)& -	& -		\\ \hline
2/5 	    & -0.43298(3) 	& -0.43277(2) 	&-0.43287(2)	& -0.43935(1) & -0.43839(2)  	&-0.43902(3)	&      0.0127(1)& 0.0113(1) 	& 0.0123(1)	\\ \hline
2/5 (with LL mixing)	    & -0.46870(10) 	& -		&		& -0.47932(17) & -	   	&-		&      0.0212(5)& -	& -		\\ \hline
2/7 	    & -0.38185(9) 	& -0.38140(6) 	&		& -0.38249(8) & -0.38188(6)  	&-		&      0.0013(3)& 0.0010(2)	& -		\\ \hline
2/9 	    & -0.34314(18) 	& -0.34274(2) 	&-0.34299(6)	& -0.34319(25)& -0.34221(2)  	&-0.34256(2)	&      0.0001(9)&     - 	& -		\\ \hline
2/11 	    & -		 	& -0.31331(2) 	&               & -	      & -0.31329(2)  	&          	&      -	& -0.0002(3)    & -		\\ \hline
2/13 	    & -		 	& -0.29087(1) 	&-0.29133(16)   & -	      & -0.29019(1)  	&-0.29047(3)	&      -	&               & -		\\ \hline
\end{tabular}
\end{center}
\caption {Lowest Landau level Coulomb interaction energies (in units of $e^2/\epsilon\ell$) obtained from a thermodynamic extrapolation of results on the spherical geometry for various spin polarized states [denoted by $(n_{\uparrow},n_{\downarrow})$] states at $\nu=2/(4p\pm1)$. The energies are obtained from exact diagonalization (``exact"), JK wave functions (``JK w.f.") or CF diagonalization (``CFD"). Also shown are the critical Zeeman energies for spin transitions between the successive states. For $E_{Z}/(e^2/\epsilon\ell)>\kappa$, the state with larger polarization is favored over the one with smaller polarization. In this as well as the following tables, the energy corrections obtained from LL mixing are quoted for the value of the LL mixing parameter $\lambda$=1.} 
\label{tab:n_2} 
\end{table*}

\begin{table*}
\begin{center}
\begin{tabular}{|c|c|c|c|c|c|c|c|c|c|}
\hline
\multicolumn{1}{|c|}{$\nu$} & \multicolumn{3}{|c|}{(3,0)} & \multicolumn{3}{|c|}{(2,1)} & \multicolumn{3}{|c|}{$\kappa$}  \\ \hline
	    & exact  		& JK w.f. 		&  CFD 		& 	exact	 & 	JK w.f. 	& CFD	 	& 	exact	& 	JK w.f. 	&  CFD		\\ \hline

3/5 	    & -0.49742(1) 	& -0.4967(3)	&		& -0.50366(2)  	 & -0.4995(1)	& -		&      0.0187(1)&     0.0081(12)& -		\\ \hline
3/5 (with LL mixing)	    & -0.57710(25) 	& -	&		& -0.58873(46)  	 & -	& -		&      0.0349(14)&     & -		\\ \hline
3/7 	    & -0.44236(2) 	& -0.4423(1) 	&-0.44237(1)	& -0.44800(2) 	 & -0.44710(1)  &-0.44748(4)	&      0.0167(2)&     0.0144(1)	& 0.0154(2)	\\ \hline
3/7 (with LL mixing)	    & -0.484663(13) 	& -	&		& -0.49628(59)  	 & -	& -		&      0.0349(14)&     & -		\\ \hline
3/11 	    & -		 	& -0.3738(1)	&		& -	      	 & -0.37352(7)	& -		&      -	&     -		& -		\\ \hline
3/13 	    & -		 	& -0.34839(3) 	&-0.34844(1)    & -	      	 & -0.34771(5) 	&-0.34794(2)	&      -	&     - 	& -		\\ \hline
3/17 	    & -		 	& -0.30924(7)	&		& -	      	 & -0.30873(6)	& -		&      -	&     -		& -		\\ \hline
\end{tabular}
\end{center}
\caption{Same as in Table \ref{tab:n_2} but for states at $\nu=3/(6p\pm1)$.} 
\label{tab:n_3} 
\end{table*}

\begin{table*}
\begin{center}
\begin{tabular}{|c|c|c|c|c|c|c|c|c|c|c|}
\hline
\multicolumn{1}{|c|}{$\nu$} & \multicolumn{3}{|c|}{(4,0)} & \multicolumn{3}{|c|}{(3,1)} & \multicolumn{3}{|c|}{(2,2)} \\ \hline
	    & exact		    & JK w.f. 	 & CFD 	       & exact		      &  JK w.f. 	  & 	CFD    & 	exact		& JK w.f. 	& CFD		\\ \hline
4/7 	    & \underbar{-0.48842(0)}& -0.4875(7) & -           &\underbar{-0.49370(0)}& -0.4904(3) & -	       &\underbar{-0.49495(0)}  & -0.4908(2)  & -    \\ \hline
4/7 (with LL mixing)	    & \underbar{-0.56161(0)}& - & -           &\underbar{-0.57177(0)}&  & -	       &\underbar{-0.57432(0)}  &   & -    \\ \hline
4/9 	    & \underbar{-0.44771(0)}& -0.44750(1)& -0.44770(10)&\underbar{-0.45241(0)}&-0.45155(1)& -0.45184(3)&\underbar{-0.45382(0)}  & -0.45275(2)& -0.45288(6) \\ \hline
4/9 (with LL mixing)	    & \underbar{-0.49303(0)}& - & -           &\underbar{-0.50256(0)}&  & -	       &\underbar{-0.50531(0)}  &   & -    \\ \hline
4/17 	    &-			    & -0.35123(1)& -0.35125(2) &- 		      &-0.35062(1)& -0.35097(6)&- 	 		& -0.35053(1)& -0.35051(1) \\ \hline
\end{tabular}
\end{center}
\caption{Same as in Table \ref{tab:n_2} but for states at $\nu=4/(8p\pm1)$. An underbar indicates that the thermodynamic extrapolation was done using only two systems, indicating that the results are less reliable.} 
\label{tab:n_4} 
\end{table*}

\begin{table*}
\begin{center}
\begin{tabular}{|c|c|c|c|c|c|c|}
\hline
\multicolumn{1}{|c|}{$\nu$} & \multicolumn{3}{|c|}{$\kappa_1$} & \multicolumn{3}{|c|}{$\kappa_2$} \\ \hline
	    & exact			& JK w.f. 		& CFD 		& exact			&  JK w.f. 		& CFD		\\ \hline
4/7 	    & \underbar{0.0211(0)}	& 0.012(4)	 	& -	 	& \underbar{0.0050(0)}	& 0.002(1)		&-		\\ \hline
4/7 (with LL mixing)	    & \underbar{0.0406(0)}	& -	 	& -	 	& \underbar{0.0102(0)}	& -		&-		\\ \hline
4/9 	    & \underbar{0.0188(0)}	&0.0162(1) 	& 0.0166(5) 	& \underbar{0.0057(0)}	&0.0048(1)	&0.0042(4)	\\ \hline
4/9 (with LL mixing)	    & \underbar{0.0381(0)}	& -	 	& -	 	& \underbar{0.0110(0)}	& -		&-		\\ \hline
\end{tabular}
\end{center}
\caption{The two critical Zeeman energies for spin transitions between the fully polarized and partially polarized states ($\kappa_1$) and between the partially polarized and spin singlet states ($\kappa_2$) at $\nu=4/(8p\pm1)$. An underbar indicates that the thermodynamic extrapolation was done using only two systems, indicating that the results are less reliable.} 
\label{tab:n_4_EZ} 
\end{table*}

\begin{table*}
\begin{center}
\begin{tabular}{|c|c|c|c|c|c|}
\hline
\multicolumn{1}{|c|}{$\nu$} & \multicolumn{1}{|c|}{(5,0)} & \multicolumn{1}{|c|}{(4,1)} & \multicolumn{1}{|c|}{(3,2)} & \multicolumn{1}{|c|}{$\kappa_1$} & \multicolumn{1}{|c|}{$\kappa_2$} \\ \hline
	    & JK w.f. 		&  JK w.f. 		& 	JK w.f. 	& JK w.f.	& JK w.f. 		\\ \hline
5/11 	    & -0.45080(1)	& -0.45429(2)	& -0.45563(4)	& 0.0175(2)	& 0.0067(3)	\\ \hline
\end{tabular}
\end{center}
\caption{Same as in Table \ref{tab:n_2} but for states at $\nu=5/(10p\pm1)$. Also shown are the two critical Zeeman energies for spin transitions between the fully polarized (5,0) and partially polarized (4,1) states ($\kappa_1$) and between the partially polarized (4,1) and partially polarized (3,2) states  ($\kappa_2$).} 
\label{tab:n_5} 
\end{table*}

\begin{table*}
\begin{center}
\begin{tabular}{|c|c|c|c|c|c|c|c|c|c|c|}
\hline
\multicolumn{1}{|c|}{$\nu$} & \multicolumn{1}{|c|}{(6,0)} & \multicolumn{1}{|c|}{(5,1)} & \multicolumn{1}{|c|}{(4,2)} & \multicolumn{2}{|c|}{(3,3)} & \multicolumn{1}{|c|}{$\kappa_1$} & \multicolumn{1}{|c|}{$\kappa_2$}  & \multicolumn{2}{|c|}{$\kappa_3$} \\ \hline
	    & JK w.f. 		&  JK w.f. 		& 	JK w.f. 	& JK w.f.	&	CFD	& JK w.f.		& JK w.f. 		& JK w.f.		& CFD$^{*}$	\\ \hline
6/13 	    & -0.45316(2)	&-0.45627(10)	& -0.45757(1)	& -0.45800(7)	& -0.45820(10)	& 0.0186(7)	& 0.0078(7)     & 0.0026(5)	& 0.0038(7)	\\ \hline
\end{tabular}
\end{center}
\caption{Same as in Table \ref{tab:n_2} but for states at $\nu=6/(12p\pm1)$. Also shown are the three critical Zeeman energies for spin transitions between the fully polarized and partially polarized (5,1) states ($\kappa_1$); between the partially polarized (5,1) and partially polarized (4,2) states  ($\kappa_2$) and between the partially polarized (4,2) and spin-singlet states  ($\kappa_3$). The $^{*}$ for the CFD value of $\kappa_3$ indicates that only the spin-singlet energy was calculated using CFD; for the partially polarized (4,2) state, the zeroth order CF energy was used.} 
\label{tab:n_6} 
\end{table*}

\section{Projections with reverse flux attachment}
\label{appendix2}

We follow the JK method of projecting a CF wave function to the lowest Landau level. We use the spherical geometry\cite{Haldane83}; $u_i=\cos(\theta_i/2)e^{i\phi_i/2}$ and $v_i=\sin(\theta_i/2)e^{-i\phi_i/2}$ are spinorial coordinates on the sphere.
First one factorizes the Jastrow factor in Eq.~\ref{Jastrow} (up to an overall sign),
\begin{eqnarray}
J^2&=&\prod_iJ_i,\\
J_i&=&\prod_{j\neq i}(u_iv_j-u_jv_i),
\end{eqnarray}
Then each $J_i$ is attached to the elements of a column of the Slater $\Phi_n$ determinant, and the projection is performed in each element individually. The Slater determinant is composed of monopole harmonics\cite{WuYang76,WuYang77} $Y_{Q,n,m}(u,v)$, where $Q$ is the effective monopole strength for composite fermions, $n$ is their $\Lambda$-level index and $m$ is the value of the $z$-component of the orbital angular momentum operator. Projection of a single electron wave function turns $Y_{Q,n,m}(u_i,v_i)$ into an operator $\hat Y_{Q,n,m}(u_i,v_i)$ that acts on the corresponding factor $J_i$.

If $2p$ flux quanta are bound to each electron, $p>1$, there one can follow two approaches. First, by Eqs.~(\ref{diff-proj_paral}) and (\ref{diff-proj_rev}), $2p-2$ powers of $J$ are moved outside of the LLL projection. Then  $\hat Y_{Q,n,m}$ acts on $J_i$. Second, if by Eqs.~(\ref{paral-flux}) and (\ref{rev-flux}) the complete Jastrow factor is within the scope of $\mathcal P_\text{LLL}$, $\hat Y_{Q,n,m}$ acts on $J^p_i$. For parallel flux attachment, the projected wave functions given in Refs.~\onlinecite{Jain97}, \onlinecite{Jain97b} and \onlinecite{Jain07} are applicable in both approaches. For reverse flux attachment ($Q<0$), Davenport and Simon\cite{Davenport12} gave an efficient method to obtain the projected wave functions. Because they implemented it explicitly only for $p=1$, here we give, for completeness, LLL projection details for $p=2$ for $Q<0$.

For $p=2$ and $Q<0$, we have
\begin{multline}
\label{proj2}
\hat Y_{Q,n,m}(u_i,v_i)J^2_i \propto \sum_{s=0}^n(-1)^s\begin{pmatrix} n \\ s\end{pmatrix}
\begin{pmatrix} 2|Q|+n \\ |Q|+m+s\end{pmatrix}\\
u_i^sv_i^{n-s}
(\partial_{u_i})^{|Q|+m+s}(\partial_{v_i})^{|Q|-m+n-s}J_i^2
\end{multline}
On the other hand,
\begin{equation}
J_i=\left(\prod_{j\neq i}\right)\sum_{t=0}^{N-1}(-1)^te_t^iv_i^{N-1-t}u_i^t,
\end{equation}
where
\begin{gather}
e_t^i=e_{t,N-1}\left(\frac{v_1}{u_1},
\dots,\frac{v_{i-1}}{u_{i-1}},\frac{v_{i+1}}{u_{i+1}},\dots,
\frac{v_N}{u_N}\right),\\
e_{t,M}(x_1,\dots,x_M)=\left\{
\begin{array}{ll}
\sum_{0<i_1<\dots\le i_M} x_{i_1}\dots x_{i_m} & \text{if $t\le M$}\\
0 & \text{otherwise}
\end{array}
\right.
\end{gather}
Here, $e_{t,m}$ are elementary symmetric polynomials. 

Thus, when evaluating Eq.~\ref{proj2}, the key step is
\begin{multline}
\label{projderiv}
(\partial_{u_i})^{|Q|+m+s}(\partial_{v_i})^{|Q|-m+n-s}
\left(\prod_{j\neq i}u_j\right)^2\times\\
\left(\sum_{t=0}^{N-1}(-1)^te_t^iv_i^{N-1-t}u_i^t\right)
\left(\sum_{r=0}^{N-1}(-1)^re_r^iv_i^{N-1-r}u_i^r\right)\\
=\left(\prod_{j\neq i}u_j\right)^2\sum_{t=0}^{N-1}(-1)^te_t^i\sum_{r=0}^{N-1}(-1)^re_r^i\times\\
(\partial_{u_i})^{|Q|+m+s}(\partial_{v_i})^{|Q|-m+n-s}v_i^{2(N-1)-t-r}u_i^{t+r}\\
=\left(\prod_{j\neq i}u_j\right)^2\sum_{t,r=0}^{N-1}(-1)^{t+r}e_t^ie_r^i\times\\
\frac{(2(N-1)-t-r)!v_i^{2(N-1)-t-r-(|Q|-m+n-s)}}{(2(N-1)-t-r-(|Q|-m+n-s))!}\times\\
\frac{(t+r)!u_i^{t+r-(|Q|+m+s)}}{(t+r-(|Q|+m+s))!}.
\end{multline}
The summation in Eq.~\ref{projderiv} must be restricted as
\begin{eqnarray}
t+r&\ge& |Q|+m+s,\\
t+r&\le& 2(N-1)-(|Q|-m+n-s).
\end{eqnarray}

The elementary symmetric polynomials stated above can be calculated iteratively using the following Newton's identity \cite{Macdonald98}:
\begin{eqnarray}
e_{m,N}(x_{1},x_{2},\cdots,x_{N})&=&\frac{1}{m}\sum_{r=1}^{m} (-1)^{r+1}p_{r,N}(x_{1},x_{2},\cdots,x_{N}) \nonumber \\ 
&&\times~ e_{m-r,N}(x_{1},x_{2},\cdots,x_{N})
\end{eqnarray}
where $p_{r,N}$ is the power-sum polynomial defined as:
\begin{equation}
p_{r,N}(x_{1},x_{2},\cdots,x_{N})=\sum_{i=1}^{N}x_{i}^{r}
\end{equation}
We also note another iterative identity of the symmetric polynomials:
\begin{multline}
e_{m,N-1}(x_{1},x_{2},\cdots,x_{j\neq i}\cdots,x_{N})=e_{m,N}(x_{1},x_{2},\cdots,x_{N}) \nonumber \\
- x_{i}e_{m-1,N-1}(x_{1},x_{2},\cdots,x_{j\neq i}\cdots,x_{N})
\end{multline}
The above two identities can be used in conjunction with each other to create an efficient routine to store the complete set of symmetric polynomials $e^{i}_{t}$. A word of caution is due here: the above quantities tend to suffer from numerical precision errors. To get around this problem, we store all the numerical values to high precision. 

We also give the projection formula for $p=1$ correcting a typo in Ref.~\onlinecite{Davenport12}:
\begin{multline}
\label{proj_p1}
\hat Y_{Q,n,m}(u_i,v_i)J_i
\propto \Big(\prod_{j\neq i}u_j\Big)\sum_{s=0}^{n}(-1)^s \binom{n}{s} \binom{2|Q|+n}{|Q|+m+s} \times \\
\sum_{t=|Q|+m+s}^{N-1-(|Q|-m+n-s)}(-1)^{t}e_t^i \frac{t!}{[t-(|Q|+m+s)]!}u_{i}^{t-(|Q|+m)} \times  \\
\frac{[N-1-t]!}{[N-1-t-(|Q|-m+n-s)]!}v_{i}^{N-1-t-(|Q|-m)}
\end{multline}

\bibliography{biblio_fqhe}

\bibliographystyle{apsrev}

\end{document}